\documentclass[twocolumn,english,prb,aps,fleqn,eqsecnum,amsmath,amssymb,floats,secnumarabic,floatfix]{revtex4}
\usepackage[T1]{fontenc}
\usepackage[latin9]{inputenc}
\usepackage{color}
\usepackage{babel}
\usepackage{units}
\usepackage{amsmath}
\usepackage{amssymb}
\usepackage{graphicx}
\usepackage{esint}
\usepackage[unicode=true,
 bookmarks=true,bookmarksnumbered=false,bookmarksopen=false,
 breaklinks=true,pdfborder={0 0 0},backref=false,colorlinks=true]
 {hyperref}
\hypersetup{pdftitle={The Hubbard model beyond the two-pole approximation: a Composite Operator Method study},
 pdfauthor={Adolfo Avella}}

\makeatletter

\providecommand{\tabularnewline}{\\}

\@ifundefined{textcolor}{}
{%
 \definecolor{BLACK}{gray}{0}
 \definecolor{WHITE}{gray}{1}
 \definecolor{RED}{rgb}{1,0,0}
 \definecolor{GREEN}{rgb}{0,1,0}
 \definecolor{BLUE}{rgb}{0,0,1}
 \definecolor{CYAN}{cmyk}{1,0,0,0}
 \definecolor{MAGENTA}{cmyk}{0,1,0,0}
 \definecolor{YELLOW}{cmyk}{0,0,1,0}
}

\usepackage{url}
\DeclareMathOperator{\Tr}{Tr}

\makeatother

\begin{document}

\title{The Hubbard model beyond the two-pole approximation:\\
a Composite Operator Method study}

\author{Adolfo Avella}

\affiliation{Dipartimento di Fisica ``E.R. Caianiello'', Universit\`a degli Studi
di Salerno, I-84084 Fisciano (SA), Italy}

\affiliation{Unit\`a CNISM di Salerno, Universit\`a degli Studi di Salerno, I-84084
Fisciano (SA), Italy}

\affiliation{CNR-SPIN, UoS di Salerno, I-84084 Fisciano (SA), Italy}
\begin{abstract}
Within the framework of the Composite Operator Method, a three-pole
solution for the two-dimensional Hubbard model is presented and analyzed
in detail. In addition to the two Hubbard operators, the operatorial
basis comprises a third operator describing electronic transitions
dressed by nearest-neighbor spin fluctuations. These latter, compared
to charge and pair fluctuations, are assumed to be preeminent in the
region of model-parameter space - small doping, low temperature and
large on-site Coulomb repulsion - where one expects strong electronic
correlations to dominate the physics of the system. This assumption
and the consequent choice for the basic field, as well as the whole
analytical approximation framework, have been validated through a
comprehensive comparison with data for local and single-particle properties
obtained by different numerical methods on varying all model parameters.
The results systematically agree, both quantitatively and qualitatively,
up to coincide in many cases. Many relevant features of the model,
reflected by the numerical data, are exactly caught by the proposed
solution and, in particular, the crossover between weak and intermediate-strong
correlations as well as the shape of the occupied portion of the dispersion.
A comprehensive comparison with other $n$-pole solutions is also
reported in order to explore and possibly understand the reasons of
such good performance.
\end{abstract}
\maketitle

\section{Introduction}

Although the number of different trials to solve more or less exactly
the two-dimensional Hubbard model \cite{Hubbard_63} are countless
and increase steadily since its advent in middle 50s, to time no analytical
approximation method can be considered to have given a clear and definitive
answer to the very many relevant issues raised by this very simple
model. This latter contains only two terms, kinetic energy and local
Coulomb repulsion, that can be cast in diagonal form in the two quantum-complementary
direct and momentum spaces. This intrinsic incompatibility leads to
many unexpected and very complex features still not all known or fully
explored, and even less deeply understood. Together with the more
fundamental and theoretical interest in this model, which is universally
considered the prototypical model for strongly correlated systems,
its relevance to real materials made the Hubbard model known in the
whole solid state community and well beyond this latter. In particular,
the model has been widely used to describe the archetypical Mott-Hubbard
insulator $V_{2}O_{3}$ \cite{Imada_98} and the cuprate high-$T_{c}$
superconductors \cite{Bednorz_86,Anderson_87}. The microscopic description
of the anomalous behaviors experimentally observed in the cuprates,
mainly in the underdoped region, in almost all experimentally measurable
physical properties \cite{Alloul_89,Timusk_99,Basov_99,Orenstein_00,Damascelli_03,Shen_05,Eschrig_06,Kanigel_06,Lee_06,Valla_06,Doiron-Leyraud_07,LeBoeuf_07,Bangura_08,Hossain_08,Yelland_08,Sebastian_08,Audouard_09,Meng_09,Anzai_10,Singleton_10,Sebastian_10,Sebastian_10a,Sebastian_10b,Tranquada_10,Vishik_10,King_11,Laliberte_11,Ramshaw_11,Riggs_11,Sebastian_11,Sebastian_11a,Sebastian_11b,Vignolle_11,Sebastian_12,Sebastian_12a}
is still an open problem. Features not predicted by standard many-body
theory and in contradiction with the Fermi-liquid framework and diagrammatic
expansions, such as non-Fermi-liquid response, quantum criticality,
pseudogap formation, ill-defined Fermi surface, kinks in the electronic
dispersion, $\ldots$ , remain still unexplained or at least controversially
debated \cite{Lee_06,Tremblay_06,Sebastian_12a}. The Hubbard model
is thought to contain by construction many of the key ingredients
necessary to explain these anomalous features: strong electronic correlations,
competition between localization and itinerancy, Mott physics, and
low-energy spin excitations.

Numerical approaches \cite{Avella_13a} are fundamental for benchmarking
and fine tuning analytical theories and for establishing which are
those capable to deal with the quite complex phenomenology of the
Hubbard model. Unfortunately, numerical techniques cannot explore,
because of their limited resolution in frequency and momentum, the
most relevant regime of model parameters (small doping, low temperature
and large on-site Coulomb repulsion) where one expects strong electronic
correlations to dominate the physics of the system. As regards analytical
and semi-analytical (i.e. embedding a numerical core) theories \cite{Avella_12},
a few are definitely worth mentioning: the work of Mori \cite{Mori_65},
Hubbard \cite{Hubbard_63,Hubbard_64,Hubbard_64a}, Rowe \cite{Rowe_68},
Roth \cite{Roth_69}, Tserkovnikov \cite{Tserkovnikov_81,Tserkovnikov_81a},
the Gutzwiller approximation \cite{Gutzwiller_63,Gutzwiller_64,Gutzwiller_65,Brinkman_70,Yokoyama_87,Gulacsi_93,Bunemann_97,Dzierzawa_97,Bunemann_98,Seibold_03,Attaccalite_03,Ferrero_05a,Capello_05,Fabrizio_07,Lanata_08,Ho_08,Deng_09,Schiro_10,Tocchio_11,Lanata_12},
the slave boson method \cite{Barnes_76,Coleman_84,Kotliar_86}, the
spectral density approach \cite{Kalashnikov_69,Nolting_72}, the two-particle
self-consistent approach \cite{Tremblay_06}, the RPA and equations-of-motion
based techniques \cite{Chubukov_04,Prelovsek_05,Plakida_06}, the
dynamical mean-field theory (DMFT) \cite{Metzner_89,Georges_92,Georges_96},
the DMFT$+\Sigma$ approach \cite{Sadovskii_05,Kuchinskii_05,Kuchinskii_06}
as well as all cluster-DMFT-like theories\cite{Maier_05} (the cellular-DMFT
\cite{Kotliar_01a}, the dynamical cluster approximation \cite{Hettler_98}
and the cluster perturbation theory \cite{Senechal_00}).

We have also been developing a systematic approach, the composite
operator method (COM) \cite{Theory,Avella_11a}, to study highly correlated
systems. In the last fifteen years, COM has been applied to several
models and materials: Hubbard \cite{Hub-redux,Odashima_05}, $p$-$d$
\cite{p-d}, $t$-$t'$-$U$ \cite{ttU-redux}, extended Hubbard ($t$-$U$-$V$)
\cite{tUV-redux}, Kondo \cite{Villani_00}, Anderson \cite{Anderson-redux},
two-orbital Hubbard \cite{Plekhanov_11}, Ising \cite{Ising-redux},
$J_{1}-J_{2}$ \cite{Avella_08a,J1J2-redux}, Cuprates \cite{Cuprates-NCA,Avella_07,Avella_07a,Avella_08,Avella_09},
etc The Composite Operator Method (COM) \cite{Theory,Avella_11a}
has the advantage to be completely microscopic, exclusively analytical,
and fully self-consistent. COM recipe uses two main ingredients \cite{Theory,Avella_11a}:
\emph{composite} operators and \emph{algebra} constraints. Composite
operators are products of electronic operators and describe the new
elementary excitations appearing in the system owing to strong correlations.
According to the system under analysis \cite{Theory,Avella_11a},
one has to choose a set of composite operators as operatorial basis
and rewrite the electronic operators and the electronic Green's function
in terms of this basis. Algebra Constraints are relations among correlation
functions dictated by the non-canonical operatorial algebra closed
by the chosen operatorial basis \cite{Theory,Avella_11a}. Other ways
to obtain algebra constraints rely on the symmetries enjoined by the
Hamiltonian under study, the Ward-Takahashi identities, the hydrodynamics,
etc \cite{Theory,Avella_11a}. Algebra Constraints are used to compute
unknown correlation functions appearing in the calculations. Interactions
among the elements of the chosen operatorial basis are described by
the residual self-energy, that is, the propagator of the residual
term of the current after this latter has been projected on the chosen
operatorial basis \cite{Theory,Avella_11a}. According to the physical
properties under analysis and the range of temperatures, dopings,
and interactions you want to explore, one has to choose an approximation
to compute the residual self-energy. In the last years, we have been
using the $n-$pole Approximation \cite{Hub-redux,Odashima_05,p-d,ttU-redux,tUV-redux,Plekhanov_11,Ising-redux,Cuprates-NCA},
the Asymptotic Field Approach \cite{Villani_00,Anderson-redux} and
the Non-Crossing Approximation (NCA) \cite{Avella_07,Avella_07a,Avella_08,Avella_09}.

In this manuscript, we present an original three-pole approximate
solution for the 2D single-band Hubbard model based on the COM. The
not-standard choice of the third field in the operatorial basis, in
addition to the two Hubbard operators, is justified in detail and
validated a posteriori by the analysis of the correlations developing
in the system. The quite involved self-consistency scheme is built
step by step and the rationale behind each assumption is given and
commented at length. The results of the proposed overall approximation
scheme are successfully compared with both numerical simulations and
other $n$-pole approximations so to fully characterize the solution
and individuate strengths, weaknesses and their sources/causes. The
main characteristics and relevant features of the proposed solutions
are summarized as well as future possible improvements are discussed.
The detailed plan of the paper follows.

In Sec.\ \ref{sec:ModelsMethods}, we discuss the model and the proposed
approximation method. In particular, in Sec.\ \ref{sec:Hamiltonian},
we present the Hubbard Hamiltonian and part of the notation we will
be using all over the manuscript. In Sec.\ \ref{sec:Basis}, we motivate
the choice of the operatorial basis and give the corresponding equations
of motion. In Sec.\ \ref{sec:Current_proj}, we discuss the projection
of the currents on the chosen basis and introduce the polar approximation.
In Sec.\ \ref{sec:Green}, we derive a closed expression for the
electronic Green's function of the system and analyze the main relations
to the relevant correlation functions. In Sec.\ \ref{sec:I}, we
discuss the normalization matrix of the system and its entries. In
Sec.\ \ref{sec:m}, we report the expression of the $m$-matrix and
analyze its properties. In Sec.\ \ref{sec:self}, we discuss in detail
the self-consistency scheme at the basis of the proposed approximation
method and the Algebra Constraints characterizing it. In Sec.\ \ref{sec:Results},
we present the results of the proposed approximation scheme. In particular,
in Sec.\ \ref{sec:charac}, we compare the three-pole approximation
presented in the manuscript with other $n$-pole approximations. In
Sec.\ \ref{sec:local}, we report a comprehensive comparison with
different numerical methods for many local properties on varying all
model parameters. In Sec.\ \ref{sec:single}, we discuss the bands
and their \emph{single} and \emph{double} occupancy in order to deeper
characterize the proposed approximation scheme. In Sec.\ \ref{sec:corr},
we report spin, charge and pair correlation functions and analyze
their behavior as function of filling and on-site Coulomb repulsion.
Finally, in Sec.\ \ref{sec:Summary}, we draw the conclusions and
present a possible outlook. In App.\ \ref{app:op_proj}, we describe
in detail the operatorial-projection scheme used to get correlation
functions of fields not belonging to the chosen operatorial basis.

\section{Model and Methods\label{sec:ModelsMethods}}

\subsection{Hamiltonian\label{sec:Hamiltonian}}

The Hamiltonian of the original (single-band, nearest-neighbor-hopping,
on-site-Coulomb-repulsion) two-dimensional Hubbard model reads as
\begin{equation}
H=-4t\sum_{\mathbf{i}}c^{\dagger}\left(i\right)\cdot c^{\alpha}\left(i\right)+U\sum_{\mathbf{i}}n_{\uparrow}\left(i\right)n_{\downarrow}\left(i\right)-\mu\sum_{\mathbf{i}}n\left(i\right)\label{eq:Ham}
\end{equation}
where
\begin{equation}
c\left(i\right)=\begin{pmatrix}c_{\uparrow}\left(i\right)\\
c_{\downarrow}\left(i\right)
\end{pmatrix}\label{eq.c-spinor}
\end{equation}
is the electronic field operator in spinorial notation and Heisenberg
picture ($i=\left(\mathbf{i},t_{i}\right)$). $\cdot$ and $\otimes$
stand for the inner (scalar) and the outer products, respectively,
in spin space. Hereafter, all composite fermionic-like operators (i.e.
composed of an odd number of original electronic operators) are written
in spinorial notation, as well as all composite bosonic-like operators
(i.e. composed of an even number of original electronic operators)
are \emph{scalars} in the same notation. $\mathbf{i}=\mathbf{R_{i}}=\left(i_{x},i_{y}\right)$
is a vector of the two-dimensional square Bravais lattice, $n_{\sigma}\left(i\right)=c_{\sigma}^{\dagger}\left(i\right)c_{\sigma}\left(i\right)$
is the particle density operator for spin $\sigma$ at site $\mathbf{i}$,
$n\left(i\right)=\sum_{\sigma}n_{\sigma}\left(i\right)=c^{\dagger}\left(i\right)\cdot c\left(i\right)$
is the total particle density operator at site $\mathbf{i}$, $\mu$
is the chemical potential, $t$ is the hopping integral and the energy
unit hereafter, $U$ is the Coulomb on-site repulsion and $\alpha_{\mathbf{ij}}$
is the projector on the nearest-neighbor sites
\begin{align}
\alpha_{\mathbf{ij}} & =\frac{1}{N}\sum_{\mathbf{k}}\mathrm{e}^{\mathrm{i}\mathbf{k}\cdot(\mathbf{R_{i}}-\mathbf{R_{j}})}\alpha\left(\mathbf{k}\right)\label{eq:alpha-ij}\\
\alpha\left(\mathbf{k}\right) & =\frac{1}{2}\left[\cos\left(k_{x}a\right)+\cos\left(k_{y}a\right)\right]\label{eq:alpha-k}
\end{align}
where $\mathbf{k}$ runs over the first Brillouin zone, $N$ is the
number of lattice sites and $a$ is the lattice constant, which will
be set to one for the sake of simplicity. For any operator $\Phi\left(i\right)$,
we use the notation $\Phi^{\kappa}\left(i\right)=\sum_{\mathbf{j}}\kappa_{\mathbf{ij}}\Phi\left(\mathbf{j},t_{i}\right)$
where $\kappa_{\mathbf{ij}}$ can be any function of the two sites
$\mathbf{i}$ and $\mathbf{j}$ and, in particular, a projector over
the cubic harmonics of the lattice: e.g. $c^{\alpha}\left(i\right)=\sum_{\mathbf{j}}\alpha_{\mathbf{ij}}c\left(\mathbf{j},t_{i}\right)$.

\subsection{Basis and equations of motion\label{sec:Basis}}

Following COM prescription \cite{Theory,Avella_11a}, we have chosen
a basic field and, in particular, we have selected the following composite
triplet field operator
\begin{equation}
\psi\left(i\right)=\begin{pmatrix}\psi_{1}\left(i\right)\\
\psi_{2}\left(i\right)\\
\psi_{3}\left(i\right)
\end{pmatrix}=\begin{pmatrix}\xi\left(i\right)\\
\eta\left(i\right)\\
c_{s}\left(i\right)
\end{pmatrix}\label{eq:basis}
\end{equation}
where $\eta\left(i\right)=n\left(i\right)c\left(i\right)$ and $\xi\left(i\right)=c\left(i\right)-\eta\left(i\right)$
are the Hubbard operators describing the electronic (charge) transitions
with filling variation per site $2\to1$ and $1\to0$, respectively.
They will give rise to the upper (UHB) and the lower (LHB) Hubbard
subbands. This choice is guided by \cite{Theory,Avella_11a}: (i)
the hierarchy of the equations of motion\cite{eta}, and (ii) by the
fact that $\xi(i)$ and $\eta(i)$ are eigenoperators\cite{eigen}
of the interacting ($U$) term of the Hamiltonian (\ref{eq:Ham}).
The fields $\xi(i)$ and $\eta(i)$ satisfy the following equations
of motion
\begin{align}
\mathrm{i}\frac{\partial}{\partial t}\xi\left(i\right) & =-\mu\xi\left(i\right)-4tc^{\alpha}\left(i\right)-4t\pi\left(i\right)\label{eq:em-xi}\\
\mathrm{i}\frac{\partial}{\partial t}\eta\left(i\right) & =\left(U-\mu\right)\eta\left(i\right)+4t\pi\left(i\right)\label{eq:em-eta}
\end{align}
where the higher-order composite field $\pi\left(i\right)$ is defined
as
\begin{equation}
\pi\left(i\right)=\frac{1}{2}n_{\mu}(i)\sigma^{\mu}\cdot c^{\alpha}\left(i\right)+c^{\dagger\alpha}\left(i\right)\cdot c\left(i\right)\otimes c\left(i\right)\label{eq:pi}
\end{equation}
and $n_{\mu}\left(i\right)=c^{\dagger}\left(i\right)\cdot\sigma_{\mu}\cdot c\left(i\right)$
is the charge- ($\mu=0$) and spin- ($\mu=1,2,3=k$) density operator,
$\sigma_{\mu}=\left(1,\vec{\sigma}\right)$, $\sigma^{\mu}=\left(-1,\vec{\sigma}\right)$,
$\sigma_{k}$ with $\left(k=1,2,3\right)$ are the Pauli matrices.

The third operator in the basis, $c_{s}\left(i\right)$, is chosen
proportional to the \emph{spin} component of $\pi\left(i\right)$:
$c_{s}\left(i\right)=n_{k}\left(i\right)\sigma_{k}\cdot c^{\alpha}\left(i\right)$.
Accordingly, we define $\bar{\pi}\left(i\right)=\pi\left(i\right)-\frac{1}{2}c_{s}\left(i\right)$.
The possibility to choose $c_{s}\left(i\right)$, or any other operator
we would consider more appropriate, instead of $\pi(i)$, which naturally
emerges from the hierarchy of the equations of motion (\ref{eq:em-xi})
and (\ref{eq:em-eta}), is a very relevant and qualifying feature
of the COM\cite{Theory,Avella_11a}. This feature makes the COM much
more flexible and effective of many other analytical approximation
techniques based on equations of motion. In particular, the use of
$c_{s}\left(i\right)$ instead of $\pi\left(i\right)$ will lead to
a great simplification in the calculations without losing, actually
highlighting, the most relevant physics. In fact, we do expect spin
fluctuations to be the most relevant fluctuations, compared to charge
and pair ones, in determining the fundamental response and the important
features of the system under analysis. We will see that this assumption
is definitely valid in the parameter regime where the electronic correlations
are expected to be very strong: large $U$, small doping $\delta=1-n$
and low temperature $T$. In absence of correlations, or for the very
weak ones, no type of fluctuations is relevant.

The field $c_{s}\left(i\right)$ satisfies the following equation
of motion
\begin{equation}
\mathrm{i}\frac{\partial}{\partial t}c_{s}\left(i\right)=-\mu c_{s}\left(i\right)+4t\kappa_{s}\left(i\right)+U\eta_{s}\left(i\right)\label{eq:em-cs}
\end{equation}
where
\begin{alignat}{1}
\kappa_{s}\left(i\right) & =\left(c^{\alpha\dagger}\left(i\right)\cdot\sigma_{k}\cdot c\left(i\right)-c^{\dagger}\left(i\right)\cdot\sigma_{k}\cdot c^{\alpha}\left(i\right)\right)\sigma_{k}\cdot c^{\alpha}\left(i\right)\nonumber \\
 & -n_{k}\left(i\right)\sigma_{k}\cdot c^{\alpha^{2}}\left(i\right)\label{eq:ks}\\
\eta_{s}\left(i\right) & =n_{k}\left(i\right)\sigma_{k}\cdot\eta^{\alpha}\left(i\right)\label{eq:etas}
\end{alignat}

\subsection{Current projection\label{sec:Current_proj}}

It is always possible to rewrite the vectorial\cite{vectorial} current
$J\left(i\right)=\mathrm{i}\frac{\partial}{\partial t}\psi\left(i\right)=\left[\psi\left(i\right),H\right]$
of the basis $\psi(i)$ as
\begin{equation}
J\left(i\right)=\sum_{\mathbf{j}}\varepsilon\left(\mathbf{i},\mathbf{j}\right)\psi\left(\mathbf{j},t\right)+\delta J\left(i\right)\label{eq:J}
\end{equation}
where the first term represents the projection of the current $J\left(i\right)$
on the basis $\psi\left(i\right)$. The proportionality matricial\cite{matricial}
function $\varepsilon\left(\mathbf{i},\mathbf{j}\right)$ is named
energy matrix: it resembles the eigenenergy for an eigenoperator of
$ $the whole Hamiltonian\cite{eigen} and it is its best approximation
for an operator that is not an eigenoperator. $\varepsilon\left(\mathbf{i},\mathbf{j}\right)$
can be computed by means of the equation
\begin{equation}
\left\langle \left\{ \delta J\left(\mathbf{i},t\right),\psi^{\dagger}\left(\mathbf{j},t\right)\right\} \right\rangle =0\label{eq:dJ}
\end{equation}
where $\left\langle \cdots\right\rangle $ stands for the thermal
average taken in the grand-canonical ensemble. The constraint (\ref{eq:dJ})
assures that the residual current $\delta J\left(i\right)$ retains/describes
only the physics \emph{orthogonal} to the one described by the chosen
basis $\psi\left(i\right)$. The constraint (\ref{eq:dJ}) gives
\begin{equation}
\varepsilon(\mathbf{k})=m(\mathbf{k})I^{-1}(\mathbf{k})\label{eq:m-I}
\end{equation}
where
\begin{equation}
\varepsilon\left(\mathbf{i},\mathbf{j}\right)=\frac{1}{N}\sum_{\mathbf{k}}\mathrm{e}^{\mathrm{i}\mathbf{k}\cdot(\mathbf{R_{i}-R_{j}})}\varepsilon\left(\mathbf{k}\right)
\end{equation}
and after having defined the normalization matrix\cite{matricial}
\begin{equation}
I\left(\mathbf{i},\mathbf{j}\right)=\left\langle \left\{ \psi\left(\mathbf{i},t\right),\psi^{\dagger}\left(\mathbf{j},t\right)\right\} \right\rangle =\frac{1}{N}\sum_{\mathbf{k}}\mathrm{e}^{\mathrm{i}\mathbf{k}\cdot(\mathbf{R_{i}-R_{j}})}I\left(\mathbf{k}\right)\label{eq:I}
\end{equation}
and the $m$-matrix\cite{matricial}
\begin{multline}
m\left(\mathbf{i},\mathbf{j}\right)=\left\langle \left\{ J\left(\mathbf{i},t\right),\psi^{\dagger}\left(\mathbf{j},t\right)\right\} \right\rangle \\
=\frac{1}{N}\sum_{\mathbf{k}}\mathrm{e}^{\mathrm{i}\mathbf{k}\cdot(\mathbf{R_{i}-R_{j}})}m\left(\mathbf{k}\right)\label{eq:m}
\end{multline}

Since $\psi\left(i\right)$ is made up of composite operators, the
normalization matrix $I\left(\mathbf{k}\right)$ is not the identity
matrix as it happens for the original electronic field operator. $I\left(\mathbf{k}\right)$
defines the spectral content of the excitations, as a function of
the momentum $\mathbf{k}$, across the band dispersion. In fact, COM\cite{Theory,Avella_11a}
has the advantage of easily and expressively describing crossover
phenomena through the transfer of weight among composite operators.

Hereafter, we will use the very convenient notation $I_{\phi\varphi}\left(\mathbf{i},\mathbf{j}\right)=\left\langle \left\{ \phi\left(\mathbf{i},t\right),\varphi^{\dagger}\left(\mathbf{j},t\right)\right\} \right\rangle $,
which generalizes the definition of the normalization matrix ($I\left(\mathbf{i},\mathbf{j}\right)=I_{\psi\psi}\left(\mathbf{i},\mathbf{j}\right)$)
and of the $m$-matrix ($m\left(\mathbf{i},\mathbf{j}\right)=I_{J\psi}\left(\mathbf{i},\mathbf{j}\right)$)
and provide the operator space of a scalar product.

\subsection{Green's functions\label{sec:Green}}

By using the decomposition of the source (\ref{eq:J}) and neglecting
the residual current $\delta J(i)$, that is, working in the framework
of a (three-)pole approximation, the retarded thermodynamic matricial\cite{matricial}
Green's functions
\begin{equation}
G\left(i,j\right)=\left\langle \mathcal{R}\left[\psi\left(i\right)\psi^{\dagger}\left(j\right)\right]\right\rangle =\theta\left(t_{i}-t_{j}\right)\left\langle \left\{ \psi\left(i\right),\psi^{\dagger}\left(j\right)\right\} \right\rangle \label{eq.G}
\end{equation}
satisfies the equation
\begin{equation}
\left(\mathrm{i}\frac{\partial}{\partial t_{i}}-\varepsilon\left(-\mathrm{i}\mathbf{\nabla}_{\mathbf{i}}\right)\right)G\left(i,j\right)=\mathrm{i}\delta\left(t_{i}-t_{j}\right)I\left(i,j\right)\label{eq:em-G}
\end{equation}

By introducing the Fourier transform $\mathcal{F}_{\mathbf{k}\omega}\left[\cdots\right]$
\begin{align}
G\left(i,j\right) & =\mathcal{F}_{\mathbf{k}\omega}\left[G\left(\mathbf{k},\omega\right)\right]\label{eq:FT-G}\\
 & =\frac{1}{N}\sum_{\mathbf{k}}\frac{\mathrm{i}}{2\pi}\int d\omega\mathrm{e}^{\mathrm{i}\mathbf{k}\cdot(\mathbf{R_{i}}-\mathbf{R_{j}})-\mathrm{i}\omega(t_{i}-t_{j})}G\left(\mathbf{k},\omega\right)
\end{align}
the equation (\ref{eq:em-G}) can be exactly solved in the frequency-momentum
space and gives
\begin{equation}
G\left(\mathbf{k},\omega\right)=\frac{1}{\omega-\varepsilon\left(\mathbf{k}\right)+\mathrm{i}\delta}I\left(\mathbf{k}\right)=\sum_{m=1}^{3}\frac{\sigma^{\left(m\right)}\left(\mathbf{k}\right)}{\omega-E_{m}\left(\mathbf{k}\right)+\mathrm{i}\delta}\label{eq:Gk}
\end{equation}
where $E_{m}\left(\mathbf{k}\right)$ are the eigenvalues of the energy
matrix $\varepsilon(\mathbf{k})$ and, as poles of the Green's function,
serve as main excitation bands of the system. $\sigma^{\left(m\right)}\left(\mathbf{k}\right)$
are the matricial\cite{matricial} spectral density weights per band
and can be computed as
\begin{equation}
\sigma_{ab}^{(m)}\left(\mathbf{k}\right)=\sum\limits _{c=1}^{3}\Omega_{am}\left(\mathbf{k}\right)\Omega_{mc}^{-1}\left(\mathbf{k}\right)I_{cb}\left(\mathbf{k}\right)\label{eq:sigmak}
\end{equation}
where the matrix $\Omega\left(\mathbf{k}\right)$ contains the eigenvectors
of $\varepsilon\left(\mathbf{k}\right)$ as columns. In the paramagnetic
phase, to which we will focus our current analysis, the diagonal terms
in spin space of all matrices involved in the calculation ($\varepsilon$,
$I$, $m$, $G$, $\sigma$, $C$, $\ldots$) are identical as well
as all off-diagonal terms in spin space of the same matrices are zero.
Accordingly, the spin index has been neglected everywhere as both
spin projections give the same result. This holds true for the energy
bands $E_{m}$ too.

The electronic Green's function $G_{cc}\left(i,j\right)=\left\langle \mathcal{R}\left[c\left(i\right)c^{\dagger}\left(j\right)\right]\right\rangle =\mathcal{F}_{\mathbf{k}\omega}\left[G_{cc}\left(\mathbf{k},\omega\right)\right]$
is given by
\begin{equation}
G_{cc}\left(\mathbf{k},\omega\right)=\sum_{a,b=1}^{2}\sum_{m=1}^{3}\frac{\sigma_{ab}^{\left(m\right)}\left(\mathbf{k}\right)}{\omega-E_{m}\left(\mathbf{k}\right)+\mathrm{i}\delta}\label{eq:Gcck}
\end{equation}
and depicts a scenario with three ($m=1,2,3$) quasi-particles with
infinite lifetimes. Their dispersions ($E_{m}\left(\mathbf{k}\right)$)
and weights ($Z_{m}\left(\mathbf{k}\right)=\sum_{ab=1}^{2}\sigma_{ab}^{\left(m\right)}\left(\mathbf{k}\right)$)
combine to give an electronic self-energy with a two-pole structure.
Such a polar structure, although leading to an electronic self-energy
with a trivial imaginary part, allows to describe the opening of gaps
and the transfer (the complete loss) of spectral weights between (within)
regions in momentum as efficiently as a full-fledge complex self-energy.

After equation (\ref{eq:sigmak}), it is obvious that the following
sum rule holds
\begin{equation}
\sum\limits _{m=1}^{3}\sigma_{ab}^{(m)}\left(\mathbf{k}\right)=I_{ab}\left(\mathbf{k}\right)\label{eq:sigmakIk}
\end{equation}

The correlation functions of the fields of the basis $C_{ab}\left(i,j\right)=\langle\psi_{a}\left(i\right)\psi_{b}^{\dagger}\left(j\right)\rangle$
can be easily determined in terms of the Green's function by means
of the spectral theorem and their Fourier transforms have the general
expression
\begin{align}
C_{ab}\left(\mathbf{k},\omega\right) & =2\pi\sum\limits _{m=1}^{3}C_{ab}^{\left(m\right)}\left(\mathbf{k}\right)\delta\left(\omega-E_{m}\left(\mathbf{k}\right)\right)\label{eq:Ck}\\
C_{ab}^{\left(m\right)}\left(\mathbf{k}\right) & =\left[1-f_{\mathrm{F}}\left(E_{m}\left(\mathbf{k}\right)\right)\right]\sigma_{ab}^{\left(m\right)}\left(\mathbf{k}\right)\label{eq:Ck_m}
\end{align}
where $f_{\mathrm{F}}\left(\omega\right)=\left(\mathrm{e}^{\frac{\omega}{k_{\mathrm{B}}T}}+1\right)^{-1}$
is the Fermi function and $C_{ab}^{\left(m\right)}\left(\mathbf{k}\right)$
is the band component per momentum of the corresponding same-time
correlation function $C_{ab}\left(\mathbf{k}\right)=\sum\limits _{m=1}^{3}C_{ab}^{\left(m\right)}\left(\mathbf{k}\right)$.
It is worth noting that the complementary correlation function $\tilde{C}_{ab}\left(i,j\right)=\langle\psi_{a}^{\dagger}\left(i\right)\psi_{b}\left(j\right)\rangle$
can be easily obtained by means of the very same ingredients ($E_{m}\left(\mathbf{k}\right)$
and $\sigma^{\left(m\right)}\left(\mathbf{k}\right)$) as
\begin{align}
\tilde{C}_{ab}\left(\mathbf{k},\omega\right) & =2\pi\sum\limits _{m=1}^{3}\tilde{C}_{ab}^{\left(m\right)}\left(\mathbf{k}\right)\delta\left(\omega-E_{m}\left(\mathbf{k}\right)\right)\label{eq:Ctk}\\
\tilde{C}_{ab}^{\left(m\right)}\left(\mathbf{k}\right) & =f_{\mathrm{F}}\left(E_{m}\left(\mathbf{k}\right)\right)\sigma_{ab}^{\left(m\right)}\left(\mathbf{k}\right)\label{eq:Ctk_m}\\
\tilde{C}_{ab}\left(\mathbf{k}\right) & =\sum\limits _{m=1}^{3}\tilde{C}_{ab}^{\left(m\right)}\left(\mathbf{k}\right)\label{eq:CtkCtk_m}
\end{align}
and that the following very useful relations hold
\begin{align}
\sigma_{ab}^{\left(m\right)}\left(\mathbf{k}\right) & =\tilde{C}_{ab}^{\left(m\right)}\left(\mathbf{k}\right)+C_{ab}^{\left(m\right)}\left(\mathbf{k}\right)\label{eq:sigmakCkCtk}\\
I_{ab}\left(\mathbf{k}\right) & =\tilde{C}_{ab}\left(\mathbf{k}\right)+C_{ab}\left(\mathbf{k}\right)\label{eq:IkCkCtk}
\end{align}

\subsection{Normalization $I$ matrix\label{sec:I}}

The normalization $I\left(\mathbf{k}\right)$ matrix has the following
symmetric structure by construction
\begin{equation}
I\left(\mathbf{k}\right)=\left(\begin{array}{ccc}
I_{11}\left(\mathbf{k}\right) & I_{12}\left(\mathbf{k}\right) & I_{13}\left(\mathbf{k}\right)\\
I_{12}\left(\mathbf{k}\right) & I_{22}\left(\mathbf{k}\right) & I_{23}\left(\mathbf{k}\right)\\
I_{13}\left(\mathbf{k}\right) & I_{23}\left(\mathbf{k}\right) & I_{33}\left(\mathbf{k}\right)
\end{array}\right)\label{eq:Ik}
\end{equation}
In a paramagnetic and homogeneous system, to which we will focus our
current analysis, its entries have the following expressions
\begin{align}
I_{11}\left(\mathbf{k}\right) & =I_{11}=1-\frac{n}{2}\label{eq:I11k}\\
I_{12}\left(\mathbf{k}\right) & =0\label{eq:I12k}\\
I_{13}\left(\mathbf{k}\right) & =3C_{\xi c}^{\alpha}+\frac{3}{2}\alpha\left(\mathbf{k}\right)\chi_{s}^{\alpha}\label{eq:I13k}\\
I_{22}\left(\mathbf{k}\right) & =I_{22}=\frac{n}{2}\label{eq:I22k}\\
I_{23}\left(\mathbf{k}\right) & =3C_{\eta c}^{\alpha}-\frac{3}{2}\alpha\left(\mathbf{k}\right)\chi_{s}^{\alpha}\label{eq:I23k}\\
I_{33}\left(\mathbf{k}\right) & =4C_{c_{s}c}^{\alpha}+\frac{3}{2}C_{\eta\eta}+3\alpha\left(\mathbf{k}\right)\left(f_{s}+\frac{1}{4}C_{cc}^{\alpha}\right)\nonumber \\
 & +\frac{3}{2}\beta\left(\mathbf{k}\right)\chi_{s}^{\beta}+\frac{3}{4}\eta\left(\mathbf{k}\right)\chi_{s}^{\eta}\label{eq:I33k}
\end{align}
where $n=\left\langle n\left(i\right)\right\rangle $ is the filling,
$\chi_{s}^{\kappa}=\frac{1}{3}\left\langle n_{k}^{\kappa}\left(i\right)n_{k}\left(i\right)\right\rangle $
is the spin-spin correlation function at distances determined by the
projector $\kappa$ and $f_{s}=\frac{1}{3}\left\langle c^{\dagger}\left(i\right)\cdot\sigma_{k}\cdot c^{\alpha}\left(i\right)n_{k}^{\alpha}\left(i\right)\right\rangle $
is a higher-order (up to three different sites are involved) spin-spin
correlation function. We have also introduced the following definitions,
which is based on those related to the correlation functions of the
fields of the basis (\ref{eq:Ck}): $C_{\phi\varphi}=\left\langle \phi_{\sigma}\left(i\right)\varphi_{\sigma}^{\dagger}\left(i\right)\right\rangle $
and $C_{\phi\varphi}^{\kappa}=\left\langle \phi_{\sigma}^{\kappa}\left(i\right)\varphi_{\sigma}^{\dagger}\left(i\right)\right\rangle $,
where no summation over sigma is intended. $\beta\left(\mathbf{k}\right)$
and $\eta\left(\mathbf{k}\right)$ are the projectors onto the second-nearest-neighbor
sites along the main diagonals and the main axes of the lattice, respectively.

\subsection{$m$-matrix\label{sec:m}}

The $m\left(\mathbf{k}\right)$ matrix has the following symmetric
structure by construction
\begin{equation}
m\left(\mathbf{k}\right)=\left(\begin{array}{ccc}
m_{11}\left(\mathbf{k}\right) & m_{12}\left(\mathbf{k}\right) & m_{13}\left(\mathbf{k}\right)\\
m_{12}\left(\mathbf{k}\right) & m_{22}\left(\mathbf{k}\right) & m_{23}\left(\mathbf{k}\right)\\
m_{13}\left(\mathbf{k}\right) & m_{23}\left(\mathbf{k}\right) & m_{33}\left(\mathbf{k}\right)
\end{array}\right)\label{eq:mk}
\end{equation}
In a paramagnetic and homogeneous system, its entries have the following
expressions
\begin{align}
m_{11}\left(\mathbf{k}\right) & =-\mu I_{11}-4t\left[\Delta+\left(p+I_{11}-I_{22}\right)\alpha\left(\mathbf{k}\right)\right]\label{eq:m11k}\\
m_{12}\left(\mathbf{k}\right) & =4t\left[\Delta+\left(p-I_{22}\right)\alpha\left(\mathbf{k}\right)\right]\label{eq:m12k}\\
m_{13}\left(\mathbf{k}\right) & =-\left(\mu+4t\alpha\left(\mathbf{k}\right)\right)I_{13}\left(\mathbf{k}\right)-4t\alpha\left(\mathbf{k}\right)I_{23}\left(\mathbf{k}\right)\nonumber \\
 & -2tI_{33}\left(\mathbf{k}\right)-4tI_{\bar{\pi}c_{s}}\left(\mathbf{k}\right)\label{eq:m13k}\\
m_{22}\left(\mathbf{k}\right) & =\left(U-\mu\right)I_{22}-4t\left[\Delta+p\alpha\left(\mathbf{k}\right)\right]\label{eq:m22k}\\
m_{23}\left(\mathbf{k}\right) & =\left(U-\mu\right)I_{23}\left(\mathbf{k}\right)+2tI_{33}\left(\mathbf{k}\right)+4tI_{\bar{\pi}c_{s}}\left(\mathbf{k}\right)\label{eq:m23k}\\
m_{33}\left(\mathbf{k}\right) & =-\mu I_{33}\left(\mathbf{k}\right)+2dtI_{\kappa_{s}c_{s}^{\dagger}}\left(\mathbf{k}\right)+UI_{\eta_{s}c_{s}^{\dagger}}\left(\mathbf{k}\right)\label{eq:m33k}
\end{align}
where $\Delta=C_{\xi\xi}^{\alpha}-C_{\eta\eta}^{\alpha}$ is the difference
between upper and lower intra-Hubbard-subband contributions to the
kinetic energy and $p=\frac{1}{4}\left(\chi_{0}^{\alpha}+3\chi_{s}^{\alpha}\right)-\chi_{p}^{\alpha}$
is a combination of the nearest-neighbor charge-charge $\chi_{0}^{\alpha}=\left\langle n^{\alpha}\left(i\right)n\left(i\right)\right\rangle $,
spin-spin $\chi_{s}^{\alpha}$ and pair-pair $\chi_{p}^{\alpha}=\left\langle \left[c_{\uparrow}\left(i\right)c_{\downarrow}\left(i\right)\right]^{\alpha}c_{\downarrow}^{\dagger}\left(i\right)c_{\uparrow}^{\dagger}\left(i\right)\right\rangle $
correlation functions. It is easy to verify that $I_{\bar{\pi}c_{s}}\left(\mathbf{k}\right)=\alpha\left(\mathbf{k}\right)I_{\bar{\pi}c_{s}}^{\alpha}$,
that is, it does not contain any same-site term and does not extend
further than nearest-neighbors.

\subsection{Self-consistency and Algebra constraints\label{sec:self}}

We can avoid cumbersome and somewhat meaningless - see in the following
- calculations by restricting $I_{\kappa_{s}c_{s}^{\dagger}}\left(\mathbf{k}\right)$
and $I_{\eta_{s}c_{s}^{\dagger}}\left(\mathbf{k}\right)$ to just
the local and the nearest-neighbor terms
\begin{equation}
m_{33}\left(\mathbf{k}\right)\cong-\mu I_{33}\left(\mathbf{k}\right)+\bar{m}_{33}^{0}+\alpha\left(\mathbf{k}\right)\bar{m}_{33}^{\alpha}\label{eq:m33k_approx}
\end{equation}
and by using a couple of Algebra constraints\cite{Theory,Avella_11a}
to compute $\bar{m}_{33}^{0}$ and $\bar{m}_{33}^{\alpha}$. As a
matter of fact, given the very complicated expressions of the composite
fields involved ($c_{s},$ $\kappa_{s}$ and $\eta_{s}$), the explicit
calculations of $I_{\kappa_{s}c_{s}^{\dagger}}\left(\mathbf{k}\right)$
and $I_{\eta_{s}c_{s}^{\dagger}}\left(\mathbf{k}\right)$ - not reported
for the sake of brevity - lead to the appearance of many unknown higher-order
correlation functions. These latter are: (i) not connected to the
chosen basis: not computable in terms of correlation functions of
the basis (\ref{eq:Ck}), (ii) not present anywhere else in the calculations:
no feedback is established to and/or from other terms, and (iii) anyway
determining uniquely the values of the cubic harmonics of $m_{33}\left(\mathbf{k}\right)$:
fixing their values by any auxiliary approximate method will be equivalent
to fix the values of the cubic harmonics of $m_{33}\left(\mathbf{k}\right)$.
Accordingly, we have chosen to fix $\bar{m}_{33}^{0}$ and $\bar{m}_{33}^{\alpha}$
directly and discard higher-order cubic harmonics taking into account
the number of Algebra constraints at our disposal (see in the following).
The very same reasoning have led us to fix $I_{\bar{\pi}c_{s}}^{\alpha}$
in the very same manner. Moreover, given the overall choice of cutting
harmonics higher than the nearest-neighbor ones, for the sake of consistency,
we also neglected the $\beta\left(\mathbf{k}\right)$ and $\eta\left(\mathbf{k}\right)$
terms in $I_{33}\left(\mathbf{k}\right)$. We checked that this latter
simplification does not lead to any appreciable difference: within
the explored paramagnetic solution, $\chi_{s}^{\beta}$ and $\chi_{s}^{\eta}$
have not very significative values. Finally, it is worth noting that
the energy matrix $\varepsilon(\mathbf{k})$ contains the inverse
of the normalization matrix $I\left(\mathbf{k}\right)$. This occurrence
implies that, although one would neglect higher harmonics in the normalization
matrix $I\left(\mathbf{k}\right)$ and in the $m\left(\mathbf{k}\right)$
matrix, the energy matrix $\varepsilon(\mathbf{k})$ could anyway
contain components at all harmonics. At least, if the normalization
matrix $I\left(\mathbf{k}\right)$ is not restricted just to the same-site
term.

It is also worth emphasizing that, although it is always possible
to use approximate methods to estimate unknown correlators, a systematic
use of this latter approach might induce uncontrolled effects on the
self-consistent scheme that could be hard to estimate by a posteriori
analysis. In this context, Algebra Constraints offer a very reliable
way to fix unknown correlators as they allow the system to adjust
its internal parameters in order to satisfy algebraic relations or
symmetry requirements which are valid for any coupling and any value
of the external parameters.

By checking systematically all operatorial relations existing among
the fields of the basis, we can recognize the following Algebra Constraints
\begin{align}
C_{\xi\xi} & =1-n+D\label{eq:Cxixi}\\
C_{\eta\eta} & =\frac{n}{2}-D\label{eq:Cetaeta}\\
C_{\xi\eta} & =0\label{eq:Cxieta}\\
C_{\xi c_{s}} & =3C_{\xi c}^{\alpha}\label{eq:Cxics}\\
C_{\eta c_{s}} & =0\label{eq:Cetacs}
\end{align}
where $D=\left\langle n_{\uparrow}\left(i\right)n_{\downarrow}\left(i\right)\right\rangle $
is the double occupancy. These relations lead to the following very
relevant ones
\begin{align}
n & =2\left(1-C_{\xi\xi}-C_{\eta\eta}\right)=2\left(\tilde{C}_{\xi\xi}+\tilde{C}_{\eta\eta}\right)\label{eq:n}\\
D & =1-C_{\xi\xi}-2C_{\eta\eta}=\tilde{C}_{\eta\eta}\label{eq:D}
\end{align}

On the other hand, we can compute $\chi_{0}^{\alpha}$, $\chi_{s}^{\alpha}$,
$\chi_{p}^{\alpha}$ and $f_{s}$ by operatorial projection (see Appendix
\ref{app:op_proj}), which is equivalent to the well-established one-loop
approximation \cite{Theory,Avella_11a} for same-time correlations
functions
\begin{align}
\chi_{0}^{\alpha} & \approx n^{2}-2\frac{I_{11}\left(C_{c\eta}^{\alpha}\right)^{2}+I_{22}\left(C_{c\xi}^{\alpha}\right)^{2}}{C_{\eta\eta}}\label{eq:chia0}\\
\chi_{s}^{\alpha} & \approx-2\frac{I_{11}\left(C_{c\eta}^{\alpha}\right)^{2}+I_{22}\left(C_{c\xi}^{\alpha}\right)^{2}}{2I_{11}I_{22}-C_{\eta\eta}}\label{eq:chias}\\
\chi_{p}^{\alpha} & \approx\frac{C_{c\xi}^{\alpha}C_{\eta c}^{\alpha}}{C_{\eta\eta}}\label{eq:chiap}\\
f_{s} & \approx-\frac{1}{2}C_{c\xi}^{\alpha}-\frac{3}{4}\chi_{s}^{\alpha}\left(\frac{C_{c\xi}^{\alpha}}{I_{11}}-\frac{C_{c\eta}^{\alpha}}{I_{22}}\right)\nonumber \\
 & -2\frac{C_{c\xi}^{\alpha}}{I_{11}}\left(C_{c\xi}^{\alpha^{2}}-\frac{1}{4}C_{c\xi}\right)-2\frac{C_{c\eta}^{\alpha}}{I_{22}}\left(C_{c\eta}^{\alpha^{2}}-\frac{1}{4}C_{c\eta}\right)\label{eq:fs}
\end{align}
As a matter of fact, the energy matrix $\varepsilon(\mathbf{k})=m(\mathbf{k})I^{-1}(\mathbf{k})$
is assured to have real eigenvalues if the normalization matrix $I\left(\mathbf{k}\right)$
is semi-positive\cite{semi-positive}. This mathematical requirement
corresponds to the physical interpretation of the eigenvalues of the
normalization matrix $I\left(\mathbf{k}\right)$ as spectral weights
of the \emph{orthogonal}, according to the defined scalar product
in the operatorial space, quasi-particles describing the system under
analysis in the given polar approximation. Then, the presence of $\chi_{s}^{\alpha}$
and $f_{s}$ in the normalization matrix $I\left(\mathbf{k}\right)$
imposes a special care in evaluating their values and, in particular,
in keeping them within their physical bounds ($-1\leq\chi_{s}^{\alpha},f_{s}\leq\frac{1}{3}$).
Any minimal diversion could lead to a negative eigenvalue in the normalization
matrix $I\left(\mathbf{k}\right)$ that is both difficult to explain
physically and hard to sustain mathematically: it can easily lead
to complex eigenvalues in the energy matrix $\varepsilon(\mathbf{k})$.
Accordingly, we have decided to avoid using Algebra Constraints to
fix them, and to fix $\chi_{0}^{\alpha}$ and $\chi_{p}^{\alpha}$
for the sake of consistency. Algebra Constraints, in the attempt to
preserve the operatorial relations they stem from, can lead to values
of the unknowns slightly off their physical bounds in the spirit of
using them as mere parameters to achieve the ultimate task of satisfying
the operatorial algebra at the level of averages.

Summarizing, we can fix the unknowns $I_{\bar{\pi}c_{s}}^{\alpha}$,
$\bar{m}_{33}^{0}$, $\bar{m}_{33}^{\alpha}$, $\mu$, $\chi_{0}^{\alpha}$,
$\chi_{s}^{\alpha}$, $\chi_{p}^{\alpha}$ and $ $$f_{s}$ through
the set of equations (\ref{eq:Cxieta}), (\ref{eq:Cxics}), (\ref{eq:Cetacs}),
(\ref{eq:n}), (\ref{eq:chia0}), (\ref{eq:chias}), (\ref{eq:chiap})
and (\ref{eq:fs}).

\section{Results\label{sec:Results}}

\subsection{Characterization within n-pole framework\label{sec:charac}}

\begin{figure}
\begin{centering}
\begin{tabular}{c}
\includegraphics[width=8cm]{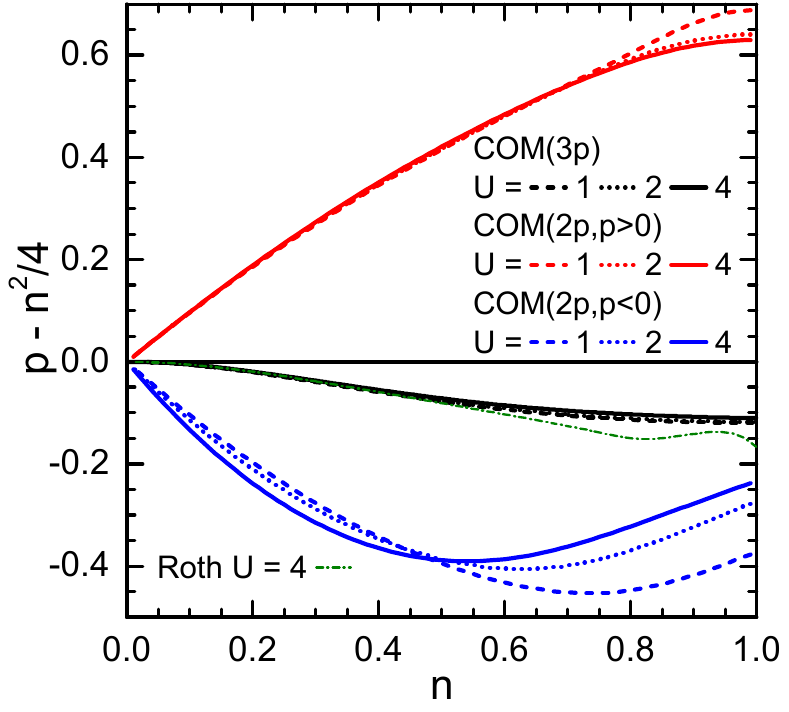}\tabularnewline
\includegraphics[width=8cm]{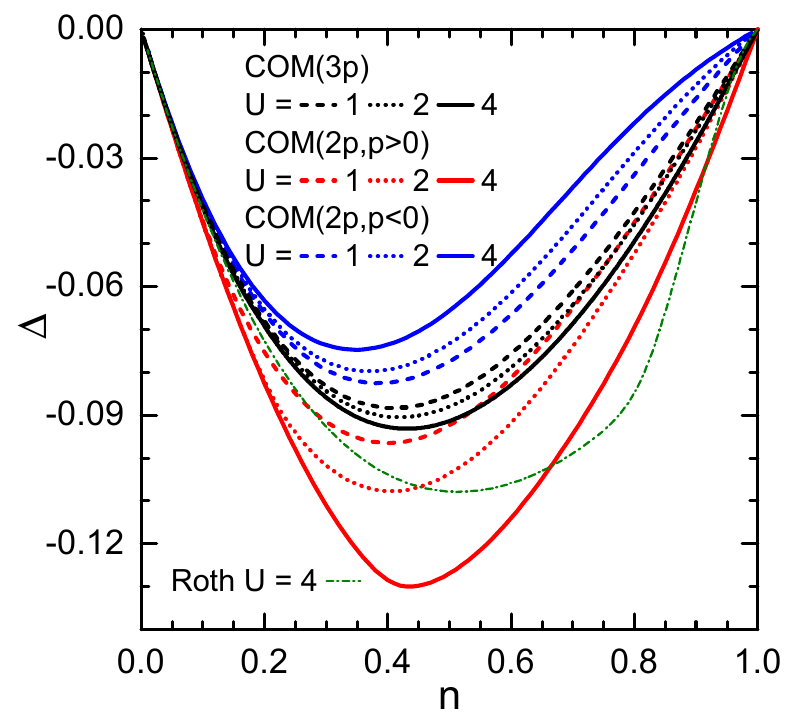}\tabularnewline
\end{tabular}
\par\end{centering}

\caption{(top) Parameter $p$, diminished of the core term $\frac{1}{4}n^{2}$
of its charge-charge correlation component $\frac{1}{4}\chi_{0}^{\alpha}$,
and (bottom) parameter $\Delta$ as functions of the filling $n$
at $U=1$ (dashed lines), $2$ (dotted lines) and $4$ (solid lines)
and $T=\nicefrac{1}{6}$ for COM(3p) (black lines), COM(2p,$p>0$)
(red lines) and COM(2p,$p<0$) (blue lines). Roth solution \cite{Roth_69}
for $U=4$ (dot-dashed green line) is also reported. \label{fig:p_Delta}}
\end{figure}

In Fig.\ \ref{fig:p_Delta} (top panel), we report the behavior of
the parameter $p$ as a function of the filling $n$ for $U=1$, $2$,
$4$ and $T=\nicefrac{1}{6}$. We subtracted the core, non-fluctuating,
term $\frac{1}{4}n^{2}$ of the charge-charge correlation component
of $p$, $\frac{1}{4}\chi_{0}^{\alpha}$, in order to be able to better
appreciate the effective intensity of the charge, spin and pair fluctuations.
The relevance of this parameter, taking into account that we will
discuss charge, spin and pair fluctuations in detail in Sec.\ \ref{sec:corr},
is strictly related to its predominant role in the characterization
of the $n$-pole solutions available in the literature. Within Hubbard
I solution, its value is approximated to just $\frac{1}{4}n^{2}$,
which corresponds to a constant value $0$ in Fig.\ \ref{fig:p_Delta}
(top panel): no charge, spin or pair fluctuations are taken into account.
The two ($p>0$ and $p<0$) two-pole COM(2p) solutions \cite{Theory,Avella_11a}
are named after the sign of $p$ as this latter completely controls
the shape of the two Hubbard subbands (see in the following) and,
consequently, the whole physical scenario underlying the dynamics
of the system. The three-pole solution COM(3p) described in the previous
section is characterized by a negative sign of $p-\frac{1}{4}n^{2}$
and by a value of this latter very similar to the one that is possible
to find by means of the Roth method \cite{Roth_69}, which is actually
based on the very same formulas (\ref{eq:chia0}), (\ref{eq:chias})
and (\ref{eq:chiap}). Although Roth uses the same formulas, the value
of $p$ and, in particular, its behavior differs quite much in the
most relevant region of filling where the effect of spin fluctuations
are expected to be more pronounced. The presence of a third field
in the basis ($c_{s}$) changes significantly the values of the correlation
functions of the basis and, consequently, those of charge, spin and
pair fluctuations. The negative sign of $p-\frac{1}{4}n^{2}$ in COM(3p)
is a clear indication of the predominance of spin fluctuations, although
with an intensity less pronounced than in COM(2p, $p<0$). Actually,
the presence of a minimum (maximum of fluctuation intensity) at a
filling significantly lower than $1$ and decreasing with increasing
$U$ in COM(2p, $p<0$) is difficult to explain as well as the so
large positive value of $p$ in COM(2p, $p>0$). As a matter of fact,
in COM(2p) solutions, the parameter $p$ is fixed by an Algebra constraint
($C_{12}=0$) and has lost its physical interpretation inherent to
its definition. Its value is just the one necessary to achieve the
fulfillment of the Pauli principle at the one-site level (i.e. $n_{\sigma}^{2}\left(i\right)=n_{\sigma}\left(i\right)$)
that is so relevant to describe correctly the spin fluctuations and,
consequently, catch the virtual processes between nearest-neighbor
sites (i.e. the scale of energy of $J=\nicefrac{4t^{2}}{U}$). In
COM(3p), $p-\frac{1}{4}n^{2}$ has just the expected behavior: it
smoothly (with respect to Roth, for instance) increases its negative
value on reducing doping.

In Fig.\ \ref{fig:p_Delta} (bottom panel), we report the behavior
of the parameter $\Delta$ as a function of the filling $n$ for $U=1$,
$2$, $4$ and $T=\nicefrac{1}{6}$. The way to fix this parameter,
together with the one used for the parameter $p$, and their resulting
values permit to characterize completely all 2-pole solutions present
in the literature. Chemical potential $\mu$, the third parameter
appearing in any 2-pole treatment, is always fixed by means of the
same equation (\ref{eq:n}) although its value and overall behavior
greatly changes according to what is used for $p$ and $\Delta$ (see
in the following). Within Hubbard I solution, the value of $\Delta$
is approximated to just $0$: no difference between the kinetic energy
contributions of the two Hubbard subbands is taken into account. While
the difference between COM solutions as regards this parameter is
not so apparent contrarily to what happens for the parameter $p$,
it is evident that Roth solution for this parameter reports a behavior
quite peculiar. Such a behavior pairs with the one of the parameter
$p$ and both cannot be easily explained and are not expected (kinks,
changes of concavity, more minima and maxima).

\subsection{Local properties and comparison with numerical results\label{sec:local}}

\begin{figure*}
\begin{centering}
\begin{tabular}{ccc}
\includegraphics[width=0.33\textwidth]{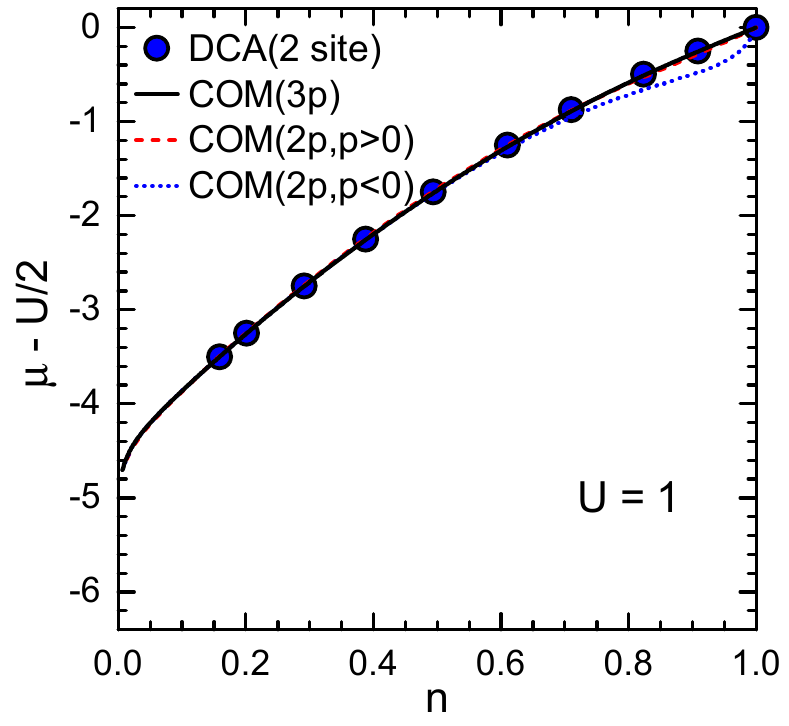} & \includegraphics[width=0.33\textwidth]{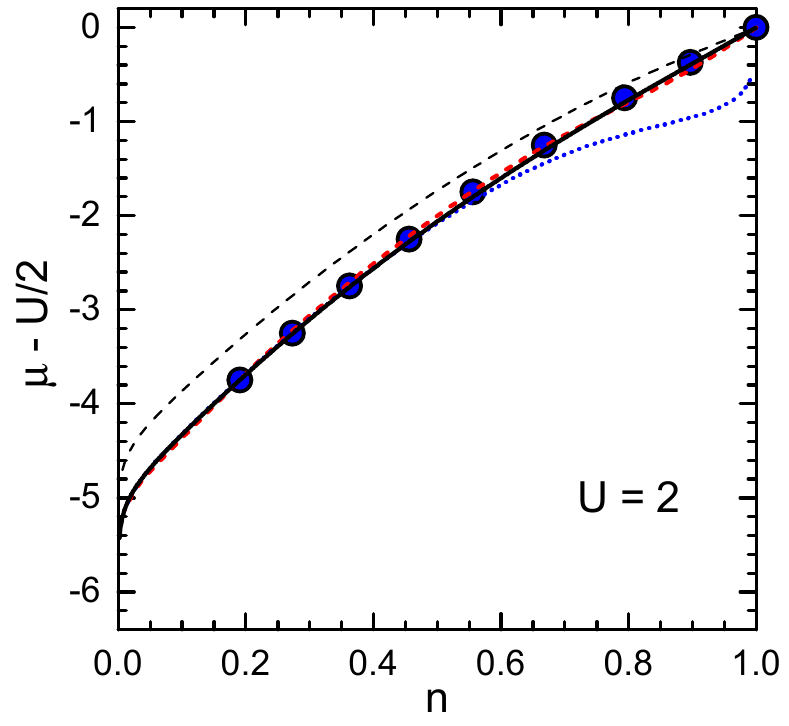} & \includegraphics[width=0.33\textwidth]{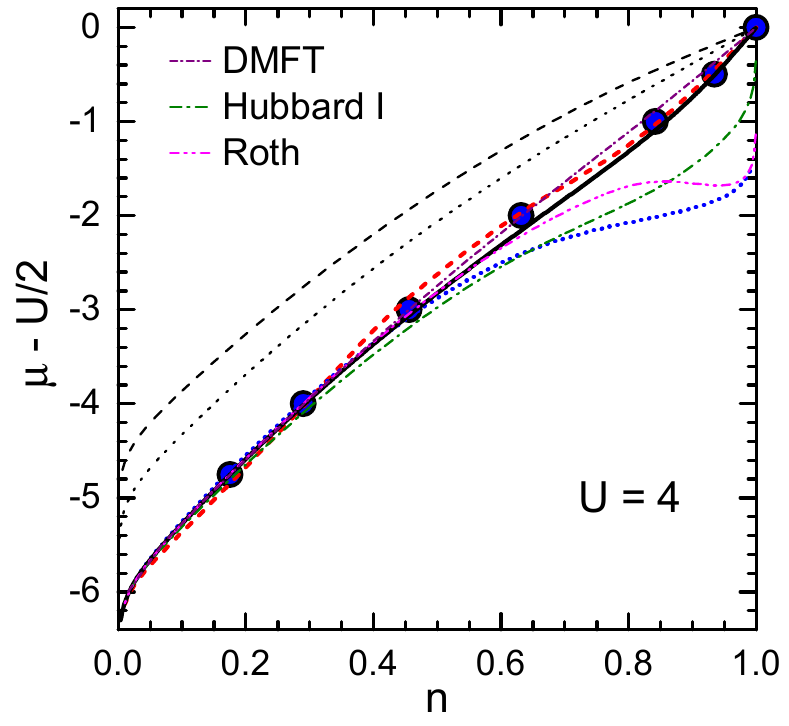}\tabularnewline
\includegraphics[width=0.33\textwidth]{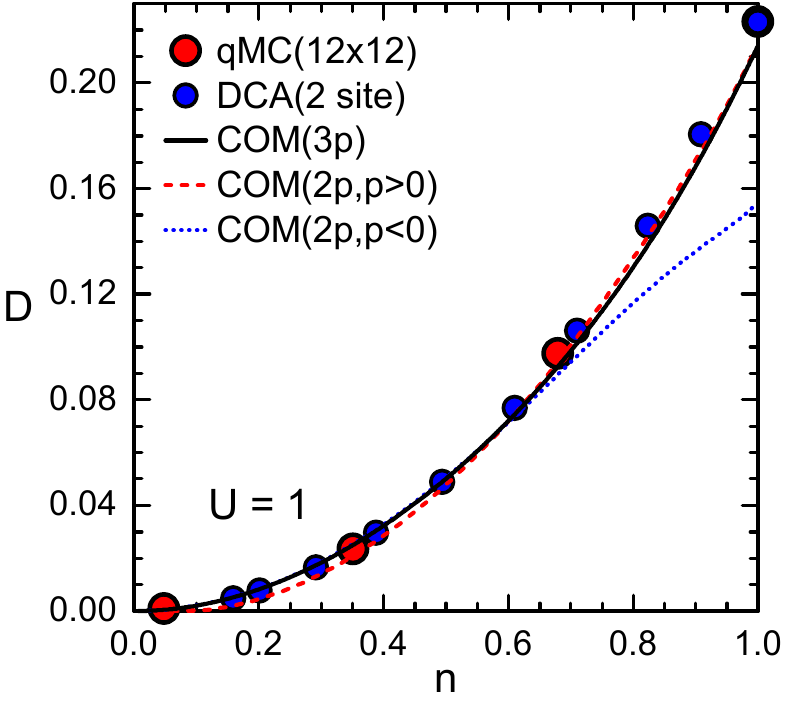} & \includegraphics[width=0.33\textwidth]{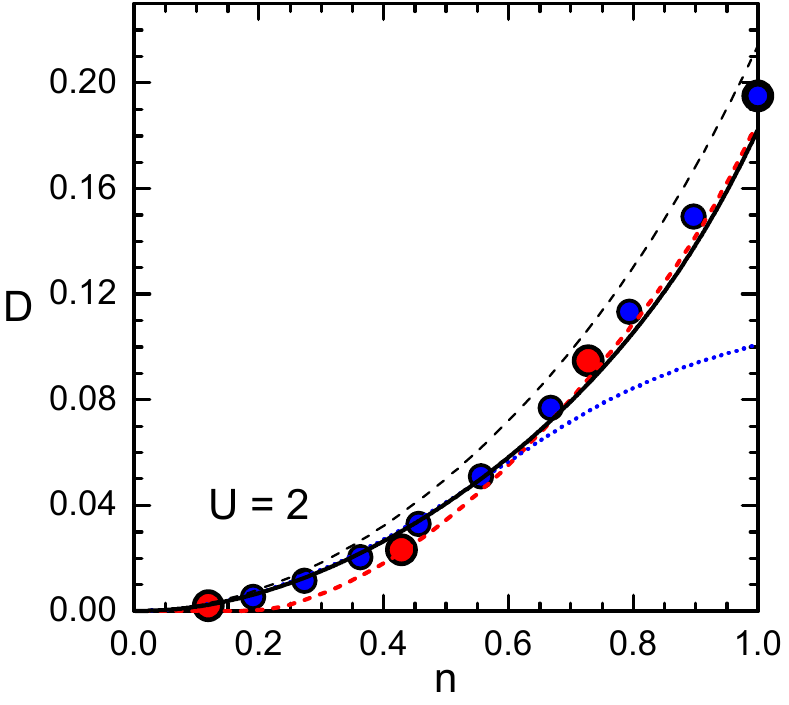} & \includegraphics[width=0.33\textwidth]{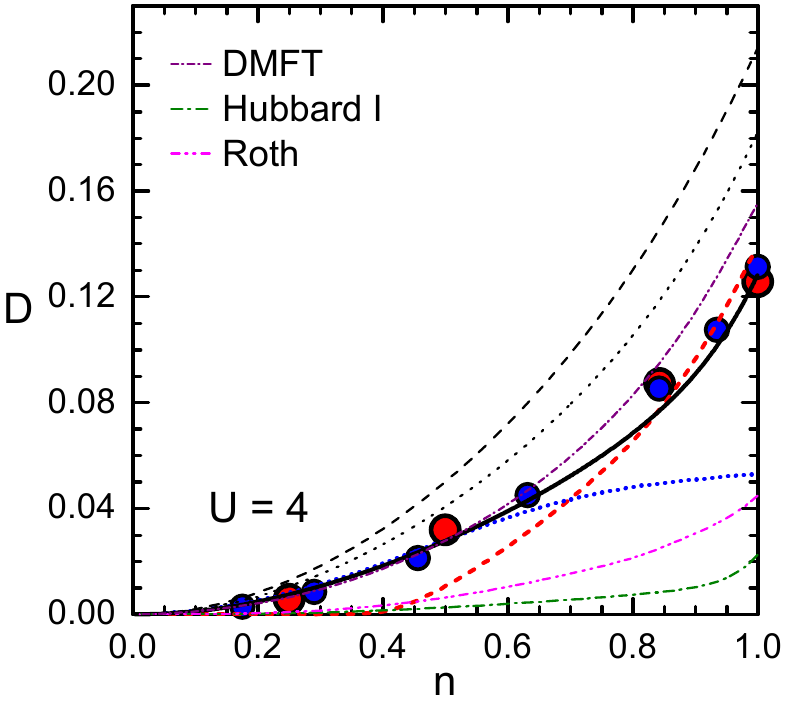}\tabularnewline
\end{tabular}
\par\end{centering}

\caption{Scaled chemical potential $\mu-\nicefrac{U}{2}$ (top row) and double
occupancy $D$ (bottom row) as functions of the filling $n$ for $U=1$
(left column), $2$ (central column) and $4$ (right column) and $T=\nicefrac{1}{6}$
for COM(3p) (black lines), COM(2p,$p>0$) (dashed red line) and COM(2p,$p<0$)
(dotted blue line). COM results are compared with $12\times12$-site
qMC \cite{Moreo_90} and $2$-site DCA \cite{Sangiovanni} numerical
data (red and blue circles, respectively) as well as with the results
of DMFT \cite{Capone} (dash-dotted purple line), Hubbard I (dot-dashed
green line) and Roth (dot-dot-dashed magenta line) methods (only at
$U=4$, right column). The thin black dashed and dotted lines in central
and right columns are COM(3p) results for $U=1$ and $U=2$, respectively.\label{fig:mu_D}}
\end{figure*}

In Fig.\ \ref{fig:mu_D}, we report the behavior of the scaled chemical
potential $\mu-\nicefrac{U}{2}$ and of the double occupancy $D$
as functions of the filling $n$ for $U=1$, $2$ and $4$ and $T=\nicefrac{1}{6}$.
It is evident the very good agreement for all values of $U$ reported
in the whole range of filling $n$ between COM(3p) and the $12\times12$-site
qMC \cite{Moreo_90} and $2$-site DCA \cite{Sangiovanni} numerical
data. The DCA data for the chemical potential show an apparent change
of concavity in proximity of half filling between $U=1,2$ and $U=4$
(Fig.\ \ref{fig:mu_D} (top row)) that is correctly caught by COM(3p)
and COM(2p, $p>0$) and not by COM(2p, $p<0$), Hubbard I and Roth
solutions, which always present the same concavity. Roth solution
actually reports a rather evident region of thermodynamic instability,
$\frac{d\mu}{dn}<0$, close to half filling. As a matter of fact,
$U=4$ induces already quite strong electronic correlations, while
$U=1,2$ do not: the chemical potential gets ready to open a gap for
higher values of $U$ and $n=1$. COM(2p, $p<0$), Hubbard I and Roth
solutions not only do not catch the change of concavity in $\mu$,
placing themselves always on the strongly correlated side, but they
also report values of $\mu$ quite far from the numerical ones: their
particle counting - actual effective filling - is definitely far from
the exact one. DMFT \cite{Capone} solution does not catch the change
of concavity for $U=4$ either (it will change concavity only for
larger values of $U$), but it features values of $\mu$ very close
to the numerical ones in the whole range of filling $n$ although
not so close as COM(3p) ones in proximity of half-filling, which is
the most interesting region. The change of correlation-strength regime
between $U=1,2$ and $U=4$ is also quite evident in the behavior
of the double occupancy $D$ (Fig.\ \ref{fig:mu_D} (bottom row)).
This latter moves from a parabolic-like behavior somewhat resembling
the non-interacting one ($\frac{n^{2}}{4}$) at $U=1,2$ (Fig.\ \ref{fig:mu_D}
(bottom-left/central panels)) to a more elaborated behavior presenting
a continuos, but well defined, change of slope on approaching half
filling at $U=4$ (Fig.\ \ref{fig:mu_D} (bottom-right panel)). Again,
COM(3p) correctly catches these features, while all other presented
solutions do not manage to achieve the same level of agreement over
the whole range of filling. Hubbard I and Roth solutions report values
of the $D$ extremely far from the numerical ones and always much
smaller than these latter, again confirming a tendency to an excess
of correlations present in such solutions. DMFT \cite{Capone} performs
extremely well, with respect to numerical data, at low-intermediate
values of filling, but at intermediate-high ones features values of
$D$ larger than the numerical ones. This is a clear evidence of a
lack of correlations for this value of $U$ as shown also by the absence
of a change in the concavity of the chemical potential. COM(2p, $p<0$)
performs really very well too at low-intermediate values of filling,
but on increasing $U$ it shows an excess of correlations close to
half filling (it is actually insulating for any finite value of $U$
at half filling - see in the following). COM(2p, $p>0$) is in very
good agreement with numerical data for $U=1$ over the whole range
of filling, but already for $U=2$, and even more for $U=4$, it is
evident a complete suppression of $D$ at low values of the filling
as well as a small, but visible, discrepancy in the slope close to
half filling for $U=4$. COM(3p) evidently has (see Fig.\ \ref{fig:mu_D}
(bottom-central/right panels)) the capability to correctly interpolate
between the two COM(2p) solutions sticking to COM(2p, $p<0$) at low-intermediate
values of filling and even improving on COM(2p, $p>0$) at intermediate-high
values of filling.

\begin{figure}
\begin{centering}
\begin{tabular}{c}
\includegraphics[width=8cm]{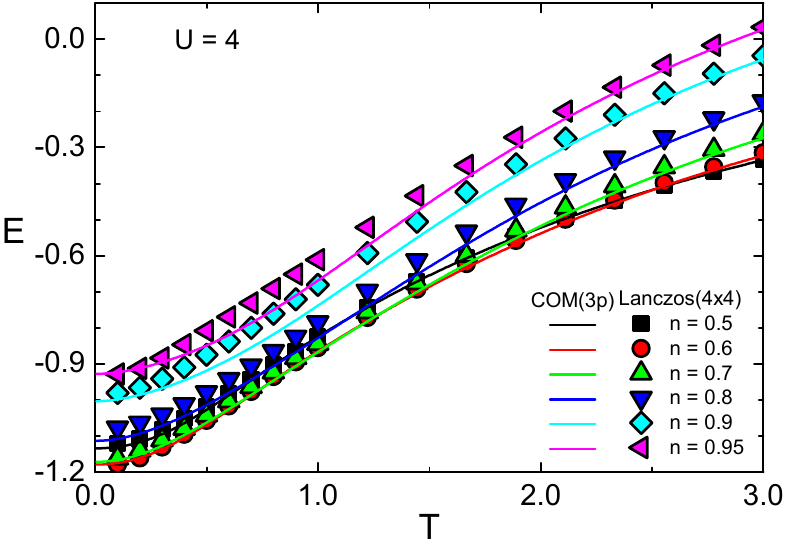}\tabularnewline
\includegraphics[width=8cm]{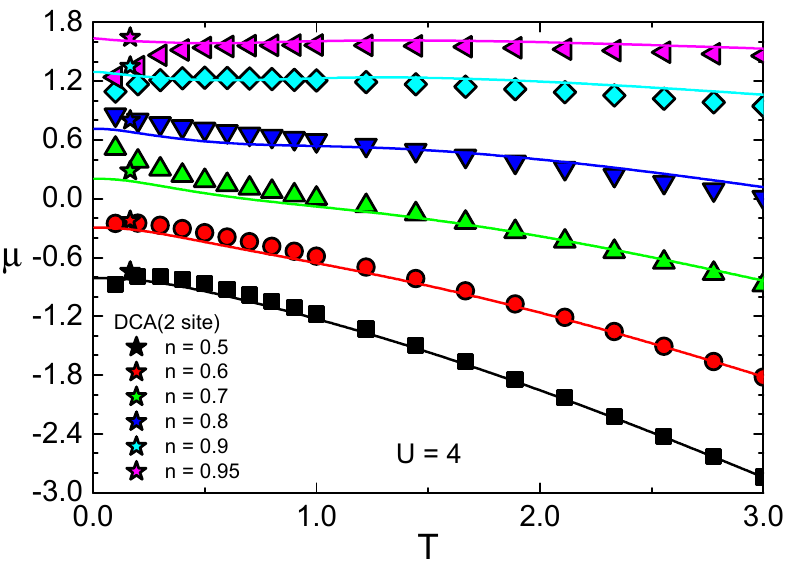}\tabularnewline
\end{tabular}
\par\end{centering}

\caption{Internal energy per site $E$ (top) and chemical potential $\mu$
(bottom) as functions of the temperature $T$ for various values of
the filling $n$ ($0.5\div0.95$) at $U=4$ for COM(3p) (solid lines),
$4\times4$-site Lanczos \cite{Prelovsek} (various symbols except
stars) and $2$-site DCA \cite{Sangiovanni} (stars) numerical data.\label{fig:D_E}}
\end{figure}

In Fig.\ \ref{fig:D_E}, we report the behavior of the internal energy
per site $E$ and of the chemical potential $\mu$ as functions of
the temperature $T$ for various values of the filling $n$ ($0.5\div0.95$)
at $U=4$. The internal energy per site $E$ has been computed as
\begin{equation}
E=\frac{1}{N}\left\langle H\right\rangle +\mu n=8tC_{cc}^{\alpha}+UD\label{eq:E}
\end{equation}
where $C_{cc}^{\alpha}=\sum_{n,m=1}^{2}C_{nm}^{\alpha}$. Given the
very good performance already discussed as regards the double occupancy,
this can be seen as a check of the capability of COM(3p) to describe
correctly the kinetic energy and, in general, the coherent transport
as a function of the temperature. As regards $E$ (Fig.\ \ref{fig:D_E}
(top panel)), the agreement between COM(3p) and Lanczos \cite{Prelovsek}
is extremely good at high temperatures, where the correlations are
weaker and Lanczos results are more reliable, and it is still really
very good at low temperatures, where, in particular for low doping,
a consistent increase of the correlations is expected. At any rate,
the very small discrepancies at low temperatures and small doping
cannot be attributed to COM(3p) as the analysis of the chemical potential
comparison will clarify. As regards the chemical potential $\mu$
(Fig.\ \ref{fig:D_E} (bottom panel)), at high temperatures the agreement
between COM(3p) and Lanczos \cite{Prelovsek} is again excellent,
but at low temperatures and for small enough doping the discrepancies
between analytical and numerical results are now very much evident
and somewhat disturbing. Now, if we add on the same graph (Fig.\ \ref{fig:D_E}
(bottom panel)) the values obtained for the reported values of the
filling by interpolating by means of cubic-splines the related DCA
results \cite{Sangiovanni} ($T=\nicefrac{1}{6}$ and $U=4$) from
Fig.\ \ref{fig:mu_D} (top row, right column), we clearly see that
DCA and Lanczos results agree very well only at high dopings (where
also COM(3p) agrees with Lanczos). At low dopings, DCA results evidently
differ from Lanczos ones and falls almost exactly on the related COM(3p)
lines. It is well known that finite-temperature Lanczos results at
low temperatures are not so reliable (they are the result of a high
temperature expansion \cite{Avella_13a}). In this specific case,
they seem to indicate the presence of a clear tendency towards a metal-insulator
transition for values of $U$ definitely too small: the chemical potential
bends towards values significantly lower than $\nicefrac{U}{2}$ for
very small doping. Such a behavior is in contrast with the more reliable
- at least in this region of model-parameter space - DCA results and
with the \emph{rehabilitated} COM(3p) ones, which, instead, very well
agrees for all values of filling and shows no tendency towards an
impending metal-insulator transition.

\begin{figure}
\begin{centering}
\begin{tabular}{c}
\tabularnewline
\end{tabular}%
\begin{tabular}{c}
\includegraphics[width=8cm]{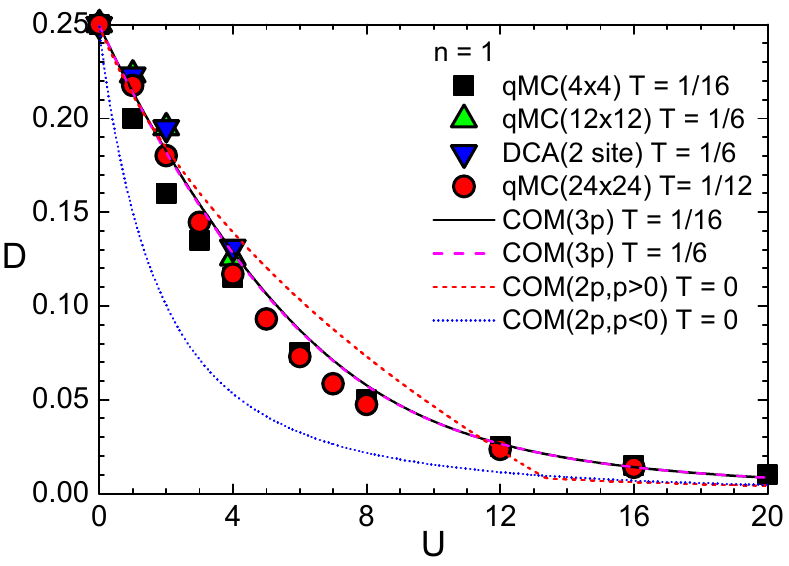}\tabularnewline
\includegraphics[width=8cm]{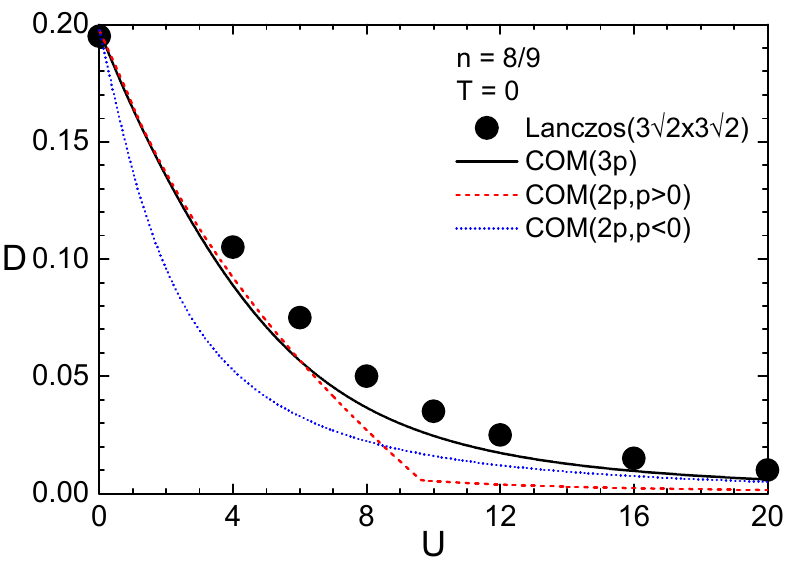}\tabularnewline
\end{tabular}
\par\end{centering}

\caption{Double occupancy $D$ as a function of the on-site Coulomb repulsion
$U$ for two values of the filling, $n=1$ (top) and $n=\nicefrac{8}{9}$
(bottom), and three values of the temperature $T$ ($0$, $\nicefrac{1}{16}$
and $\nicefrac{1}{6}$) for COM(3p) (black solid and magenta dashed
lines), COM(2p,$p>0$) (red dashed line) and COM(2p,$p<0$) (blue
dotted line). COM results are compared with $4\times4$-site qMC \cite{White_89}
(black square), $12\times12$-site qMC \cite{Moreo_90} (green up-triangle),
$2$-site DCA \cite{Sangiovanni} (blue down-triangle), $24\times24$-site
qMC \cite{Varney_09} (red circle in the upper panel) and $3\sqrt{2}\times3\sqrt{2}$-site
Lanczos \cite{Becca_00} (black circle in the lower panel) numerical
data.\label{fig:D}}
\end{figure}

\begin{figure*}
\begin{centering}
\begin{tabular}{ccc}
\includegraphics[width=0.33\textwidth]{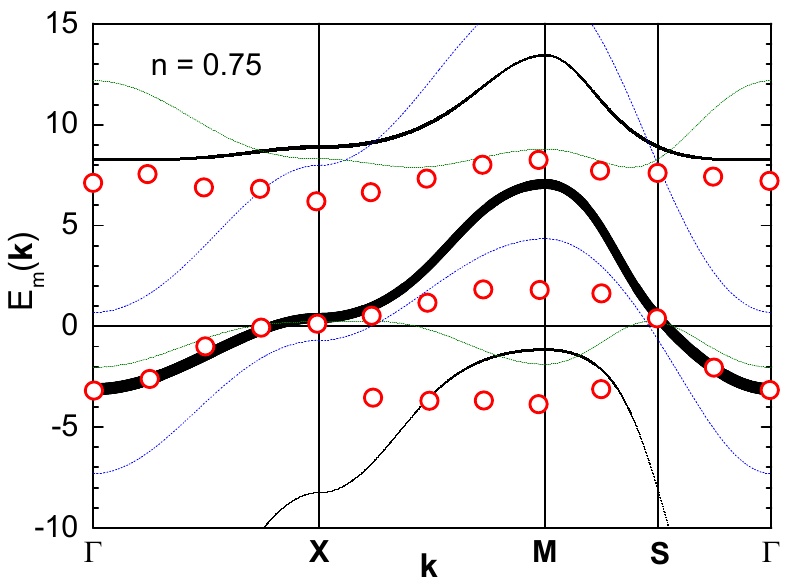} & \includegraphics[width=0.33\textwidth]{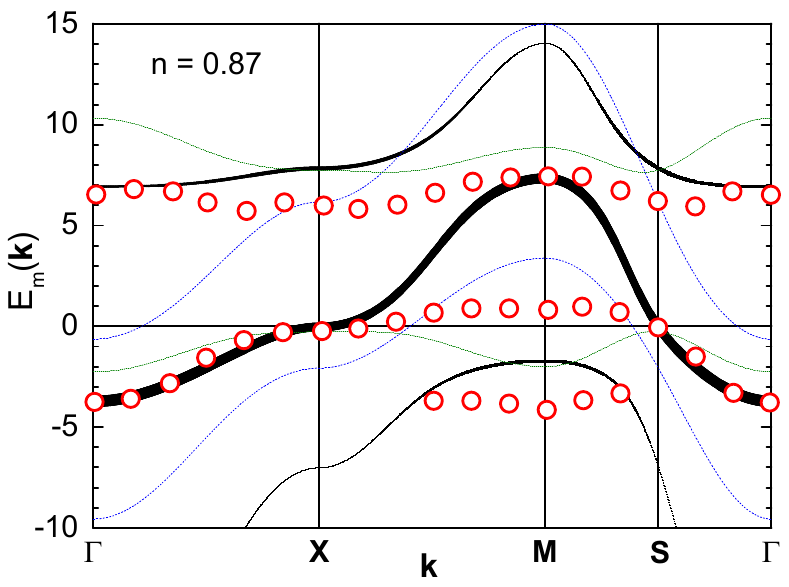} & \includegraphics[width=0.33\textwidth]{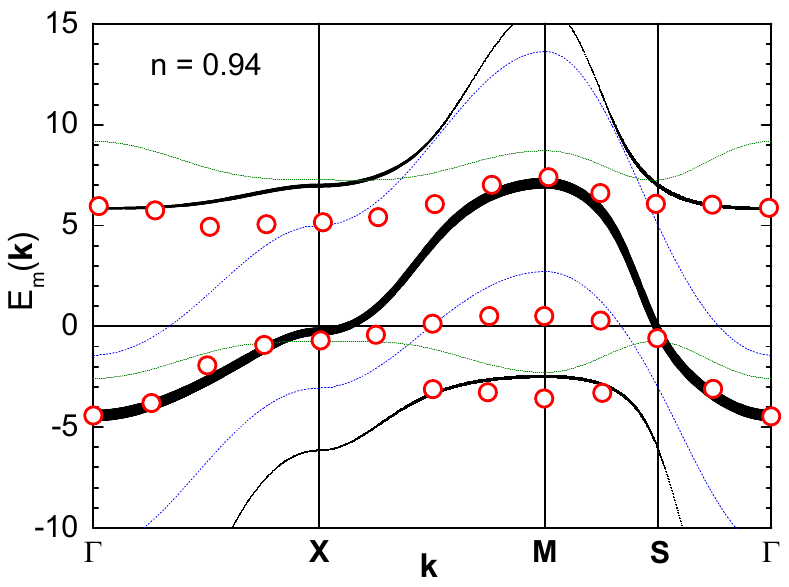}\tabularnewline
\end{tabular}
\par\end{centering}

\caption{Energy bands $E_{m}\left(\mathbf{k}\right)$ along the principal directions
of the first Brillouin zone ($\Gamma=(0,0)$ $\to$ $X=(\pi,0)$ $\to$
$M=(\pi,\pi)$ $\to$ $S=(\nicefrac{\pi}{2},\nicefrac{\pi}{2})$ $\to$
$\Gamma=(0,0)$) for $T=0.5$, $U=8$ and three different values of
the filling $n=0.75$, $0.87$ and $0.94$ for COM(3p) (variable-thickness
black line), COM(2p,$p>0$) (thin dashed blue line) and COM(2p,$p<0$)
(thin dotted green line). As regards COM(3p), the thickness of each
band is proportional to the value of the corresponding electronic
spectral density weight $\sigma_{cc}^{\left(m\right)}\left(\mathbf{k}\right)$
(see in the text). COM results are compared with qMC numerical data
\cite{Bulut_94a} (red hollow circles).\label{fig:Ek}}
\end{figure*}

In Fig.\ \ref{fig:D}, we report the behavior of the double occupancy
$D$ as a function of the on-site Coulomb repulsion $U$ for two values
of the filling $n$ ($1$ and $\nicefrac{8}{9}$) and four values
of the temperature $T$ ($0$, $\nicefrac{1}{16}$, $\nicefrac{1}{12}$
and $\nicefrac{1}{6}$). First of all, it is worth noting that all
reported COM solutions exactly reproduce, by construction, both the
$U\rightarrow0$ and the $U\rightarrow\infty$ limits. At half filling
(Fig.\ \ref{fig:D} (top panel)), COM(3p) results do not show any
appreciable difference between the two reported temperatures ($T=\nicefrac{1}{16}$
and $T=\nicefrac{1}{6}$) while numerical data show some difference
at low values of $U$, where four sets are available at the same time.
At low values of $U$, COM(3p) results agree very well with the $12\times12$-site
qMC \cite{Moreo_90}, $2$-site DCA \cite{Sangiovanni} and $24\times24$-site
qMC \cite{Varney_09} data, which are definitely more reliable of
the $4\times4$-site qMC \cite{White_89} data because of both the
numerical method used (DCA) and the size of the clusters involved
($12\times12$ and $24\times24$). At intermediate-high values of
$U$, COM(3p) results agree quite well with both $4\times4$-site
qMC \cite{White_89} and $24\times24$-site qMC \cite{Varney_09}
data, which almost exactly coincide, and agree exactly with both of
them for the higher reported values of $U$. At $n=\nicefrac{8}{9}$
(Fig.\ \ref{fig:D} (bottom panel)), the agreement between COM(3p)
and the $3\sqrt{2}\times3\sqrt{2}$-site Lanczos \cite{Becca_00}
data is quite good and improves more and more on increasing $U$.
On the other hand, the overlap of the reported numerical data at half
filling and at $n=\nicefrac{8}{9}$ already in the intermediate range
of values of $U$ (we can expect it only at sufficiently high values
of $U$) is quite suspect and calls for a revisiting by means of advanced
numerical methods applied to larger clusters. Comparing COM(3p) results
to COM(2p) ones, we immediately see that COM(3p) overcomes both (i)
the very pronounced kink characteristic of COM(2p,$p>0$), signaling
the opening of the gap at half-filling and the exit of the chemical
potential from the upper Hubbard band at $n=\nicefrac{8}{9}$, and
(ii) the exceedingly small values of $D$ at low and zero doping in
COM(2p,$p<0$) already discussed before. The opening of the gap at
half-filling and the exit of the chemical potential from the upper
Hubbard band at finite doping are strictly equivalent processes with
respect to the double occupancy $D$ in COM(2p,$p>0$) as the vast
majority of the contribution to $D$ in COM(2p,$p>0$) comes from
the upper Hubbard band - see in the following.

\subsection{Single-particle properties\label{sec:single}}

In Fig.\ \ref{fig:Ek}, we report the energy bands $E_{m}\left(\mathbf{k}\right)$
along the principal directions of the first Brillouin zone ($\Gamma=(0,0)$
$\to$ $X=(\pi,0)$ $\to$ $M=(\pi,\pi)$ $\to$ $S=(\nicefrac{\pi}{2},\nicefrac{\pi}{2})$
$\to$ $\Gamma=(0,0)$) for $T=0.5$, $U=8$ and three different values
of the filling $n=0.75$, $0.87$ and $0.94$. As regards COM(3p),
the thickness of each band is proportional to the value of the corresponding
electronic spectral density weight $\sigma_{cc}^{\left(m\right)}\left(\mathbf{k}\right)=\sum_{a,b=1}^{2}\sigma_{ab}^{\left(m\right)}\left(\mathbf{k}\right)$.
This latter corresponds to the component per band of the momentum
distribution function per spin $n\left(\mathbf{k}\right)$ at $T=0$
for those bands and momenta below the chemical potential. Such a decoration
shows the effective relevance of each energy bands, momentum per momentum,
with respect to actual occupation and possible hole/electron doping.
For all three reported values of the filling, COM(3p) results are
in very good agreement with qMC numerical data \cite{Bulut_94a} as
regards the occupied part of the central band (CB). It is worth noticing
that this is the most reliable portion of the numerical data as it
is close to the chemical potential and tracks the occupied true quasi-particle
peak. The lower band (or relic of a band) found by qMC is known as
shadow band and has very low intensity. An intensity so low as not
to allow a very precise determination of its position, given the very
significative broadening of the corresponding peaks. At any rate,
this structure is mimed by the LHB in COM(3p), which has a not-negligible
occupation right close to $M$ point. On decreasing the doping, this
correspondence becomes more and more faithful up to be almost perfect,
except for the concavity, at the lower value of the doping ($n=0.94$).
The portion of the numerical data closer to the chemical potential
at the $M$ point also has not very relevant intensity, but it is
important to describe the way the system approaches the metal-insulator
transition at half-filling. Unfortunately, this portion of the numerical
data is completely missed by COM(3p) solution. This latter also presents
a finite occupation of the CB between the two main numerical bands.
We can easily recognize that the upper numerical band (whose intensity
is significative only close to $M$ point) is very well mimed by the
UHB of COM(3p) close to $\Gamma$ point and by the CB close to $M$
point. As well as for the numerical shadow band, on decreasing the
doping, the agreement becomes better and better up to be really very
good at the lower value of the doping ($n=0.94$) close to both $\Gamma$
and $M$ points. It is worth reminding that qMC data are more and
more severely affected by the sign problem on increasing doping: high
doping results are less reliable and have larger error bars. As regards
the comparison of COM(3p) solution with COM(2p) ones, it is evident
the great number and the high level of similarities with the two COM(2p,$p<0$)
bands. These latter seem to interpolate somehow between the three
COM(3p) bands. It is worth noticing that although COM(2p,$p<0$) bands
are really very close to the numerical data in proximity of both the
$\Gamma$ point (LHB) and the $M$ point (UHB), COM(3p) bands just
lie behind the numerical points in those regions. This clearly shows
that the addition of the third field has definitely improved the overall
description of the dynamics. COM(2p,$p>0$) bands are simply too different
to make any kind of sensible comment.

\begin{figure*}
\begin{centering}
\begin{tabular}{ccc}
\includegraphics[width=0.33\textwidth]{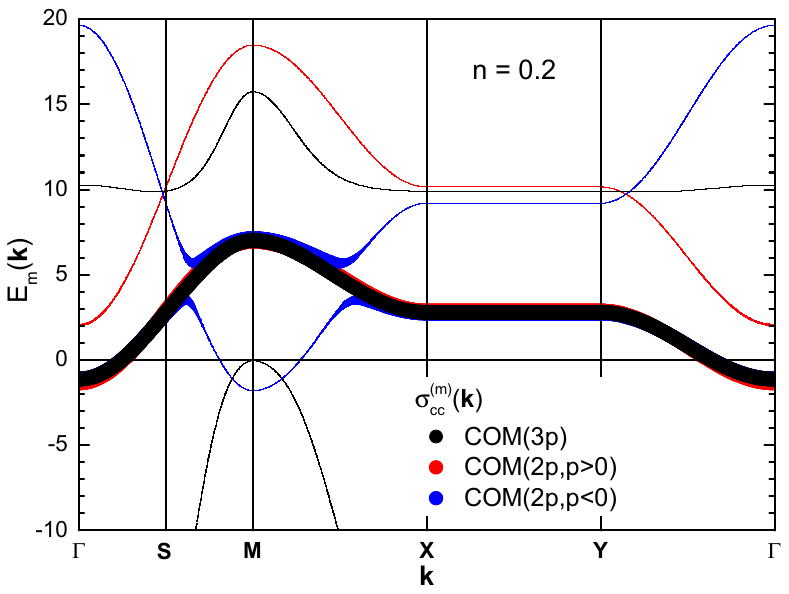} & \includegraphics[width=0.33\textwidth]{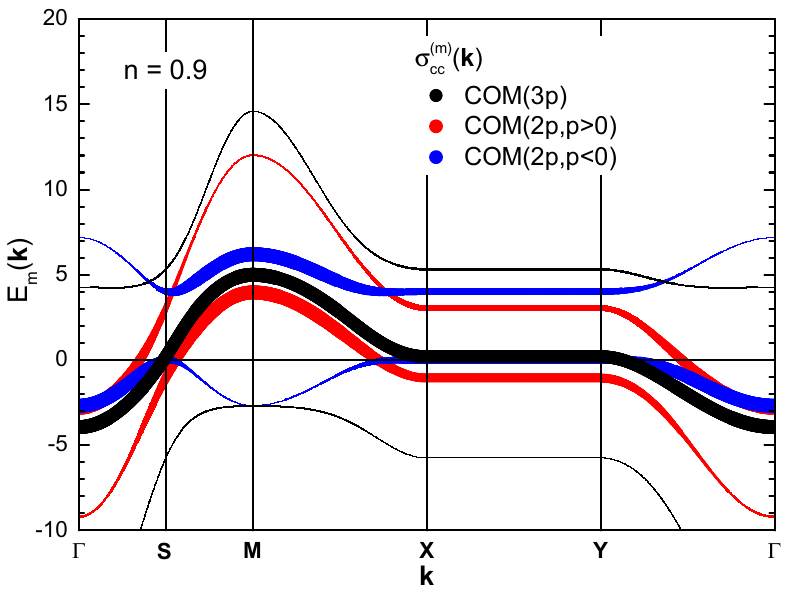} & \includegraphics[width=0.270147\textwidth]{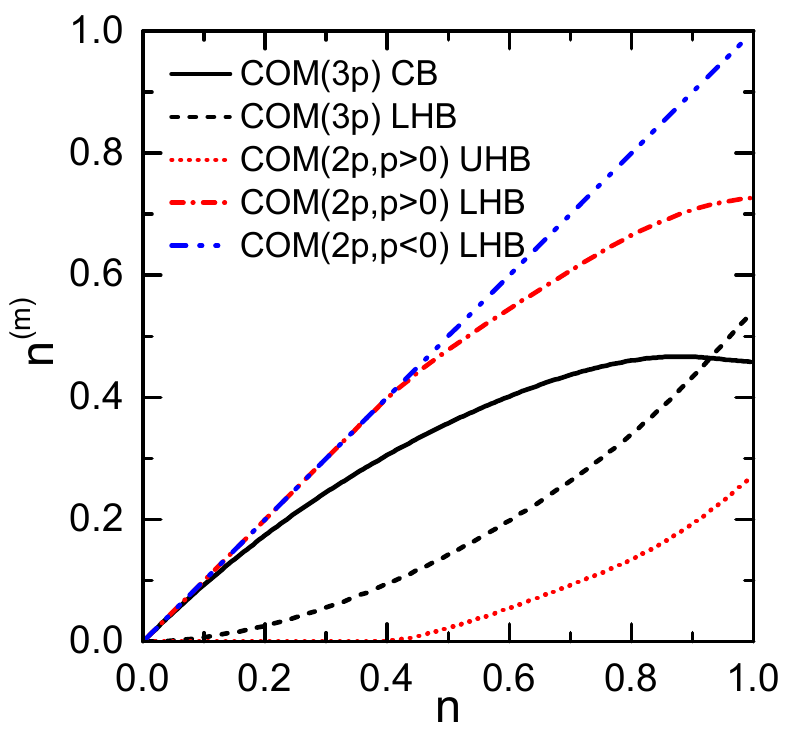}\tabularnewline
\includegraphics[width=0.33\textwidth]{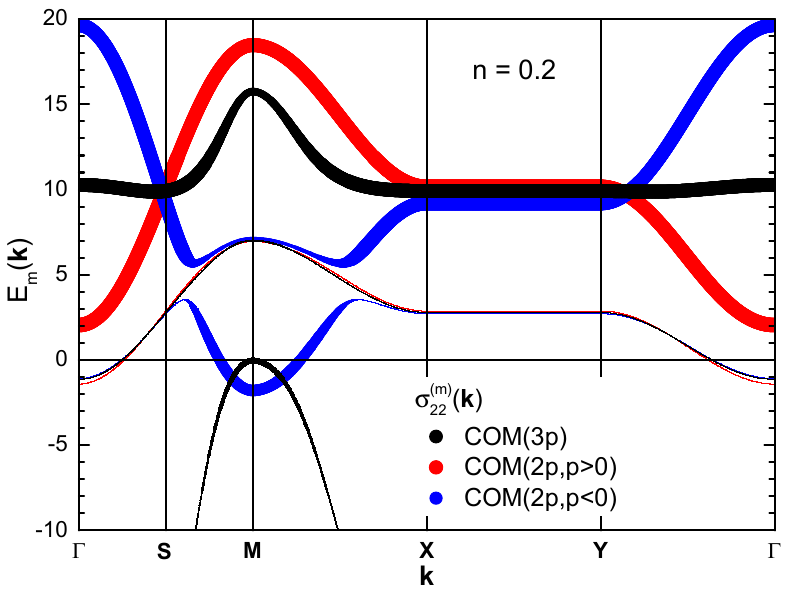} & \includegraphics[width=0.33\textwidth]{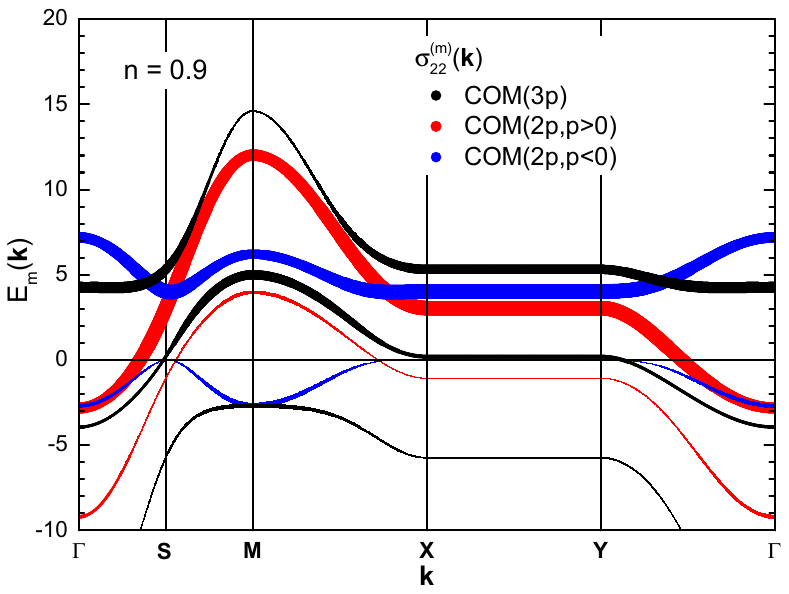} & \includegraphics[width=0.276935\textwidth]{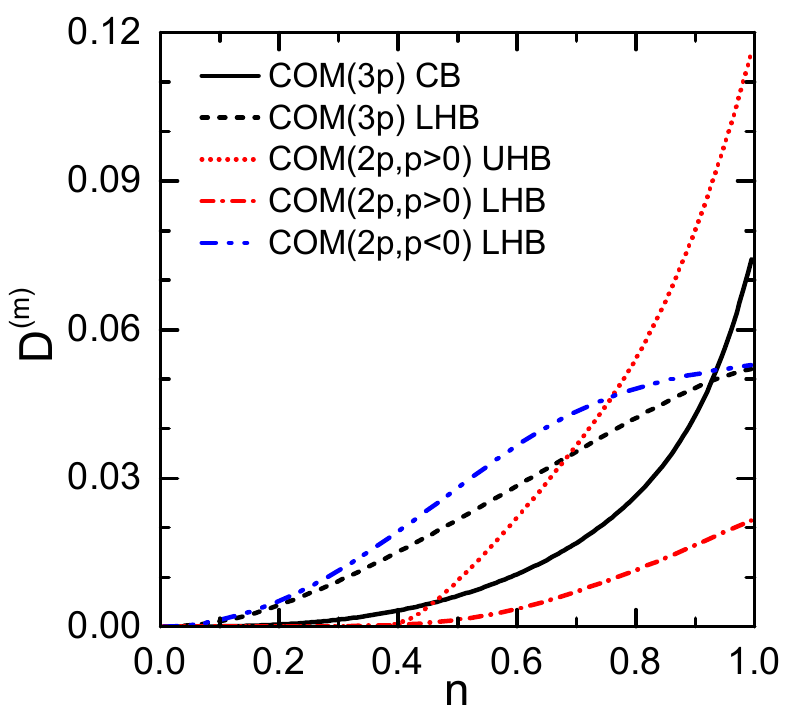}\tabularnewline
\end{tabular}
\par\end{centering}

\caption{Energy bands $E_{m}\left(\mathbf{k}\right)$ along the principal directions
of the first Brillouin zone ($\Gamma=(0,0)$ $\to$ $S=(\nicefrac{\pi}{2},\nicefrac{\pi}{2})$
$\to$ $M=(\pi,\pi)$ $\to$ $X=(\pi,0)$ $\to$ $Y=(0,\pi)$ $\to$
$\Gamma=(0,0)$) at $T=\nicefrac{1}{6}$, $U=4$ and two different
values of the filling $n=0.2$ (left column) and $n=0.9$ (central
column) for COM(3p) (variable-thickness black line), COM(2p,$p>0$)
(variable-thickness red line) and COM(2p,$p<0$) (variable-thickness
blue line). The thickness of each band is proportional to the value
of the corresponding electronic spectral density weight $\sigma_{cc}^{\left(m\right)}\left(\mathbf{k}\right)$
in the top row and to the value of the corresponding spectral density
weight $\sigma_{22}^{\left(m\right)}\left(\mathbf{k}\right)$ in the
bottom row. Band components of the filling $n^{\left(m\right)}$ (top-right)
and of the double occupancy $D^{\left(m\right)}$ (bottom-right) as
functions of the filling $n$ at $T=\nicefrac{1}{6}$ and $U=4$ for
COM(3p) (black lines: CB solid, LHB dashed), COM(2p,$p>0$) (red lines:
UHB dotted, LHB dot-dashed) and COM(2p,$p<0$) (blue dot-dot-dashed
line).\label{fig:Eksk}}
\end{figure*}

In Fig.\ \ref{fig:Eksk} (left and central columns), we report the
energy bands $E_{m}\left(\mathbf{k}\right)$ along the principal directions
of the first Brillouin zone ($\Gamma=(0,0)$ $\to$ $S=(\nicefrac{\pi}{2},\nicefrac{\pi}{2})$
$\to$ $M=(\pi,\pi)$ $\to$ $X=(\pi,0)$ $\to$ $Y=(0,\pi)$ $\to$
$\Gamma=(0,0)$) at $T=\nicefrac{1}{6}$, $U=4$ and two different
values of the filling $n=0.2$ and $n=0.9$. The thickness of each
band is proportional to the value of the corresponding electronic
spectral density weight $\sigma_{cc}^{\left(m\right)}\left(\mathbf{k}\right)$
in the top row and $\sigma_{22}^{\left(m\right)}\left(\mathbf{k}\right)$
in the bottom row. After (\ref{eq:n}), (\ref{eq:D}), (\ref{eq:Ctk_m})
and (\ref{eq:CtkCtk_m}), they are the component per band and momentum
of the filling $n$ and of double occupancy $D$, respectively, at
$T=0$ for those bands and momenta below the chemical potential:
\begin{align}
n & =\sum_{m=1}^{3}n^{\left(m\right)}=\sum\limits _{m=1}^{3}\left[\frac{1}{N}\sum_{\mathbf{k}}f_{\mathrm{F}}\left(E_{m}\left(\mathbf{k}\right)\right)\sigma_{cc}^{\left(m\right)}\left(\mathbf{k}\right)\right]\label{eq:nm}\\
D & =\sum_{m=1}^{3}D^{\left(m\right)}=\sum\limits _{m=1}^{3}\left[\frac{1}{N}\sum_{\mathbf{k}}f_{\mathrm{F}}\left(E_{m}\left(\mathbf{k}\right)\right)\sigma_{22}^{\left(m\right)}\left(\mathbf{k}\right)\right]\label{eq:Dm}
\end{align}
In Fig.\ \ref{fig:Eksk} (right column), we report the band component
of the filling $n^{\left(m\right)}$ (top-right panel) and of the
double occupancy $D^{\left(m\right)}$ (bottom-right panel) as functions
of the filling $n$ for the same values of temperature ($T=\nicefrac{1}{6}$)
and on-site Coulomb repulsion ($U=4$).

At $n=0.2$, we expect a significative reduction of the correlations
given that the average distance between particles is greater than
$2$ lattice spacings. For this filling, it is evident that the bands
collecting the vast majority of the electronic occupancy (Fig.\ \ref{fig:Eksk}
(top-left panel)) are almost identical across all reported COM solutions.
Looking at $n^{\left(m\right)}$ (Fig.\ \ref{fig:Eksk} (top-right
panel)) for the same value of filling, we immediately realize that
actually COM(3p) is characterized by a small, but finite, occupation
of its LHB, besides the occupation of its CB, which is the band coinciding
with the COM(2p) LHBs. This can be understood in terms of the proximity
of COM(3p) LHB to the chemical potential at the $M$ point. There,
COM(3p) LHB also features a maximum, that is, a very high density
of states (see in the following). Looking instead at $D^{\left(m\right)}$
(Fig.\ \ref{fig:Eksk} (bottom-right panel)) at $n=0.2$ and cross
checking with the $\sigma_{22}^{\left(m\right)}\left(\mathbf{k}\right)$
spectral density (Fig.\ \ref{fig:Eksk} (bottom-left panel)), we
can understand why COM(2p,$p>0$) features a vanishing double occupancy
at small fillings (see Fig.\ \ref{fig:mu_D} (bottom-central/right
panels)). Double occupancy is negligible in COM(2p,$p>0$) LHB and
all concentrated in the UHB that is all above the chemical potential
at small fillings. COM(2p,$p<0$) and COM(3p) LHBs instead contribute
significatively to the actual value of the double occupancy at all
values of the filling. They contribute almost identically at small
and large fillings and very much similarly at intermediate fillings
although the actual shape of the bands is quite different away from
the $M$ point. As a matter of fact, LHB is the only occupied band
in COM(2p,$p<0$) (Fig.\ \ref{fig:Eksk} (top-right panel)) at all
finite values of $U$ (see in the following) and this is the reason
why the double occupancy is so small at intermediate and large fillings.
Contrarily, at large fillings, COM(3p) can count on the contribution
of its CB to the double occupancy, which is greatly enhanced by the
proximity of the van Hove singularity to the chemical potential. The
composition of these two contributions to the double occupancy (COM(3p)
LHB and CB ones) and their quite different behavior with filling (Fig.\ \ref{fig:Eksk}
(bottom-right panel)) can explain the evident change of slope on approaching
half filling (see Fig.\ \ref{fig:mu_D} (bottom row, right column)).

At $n=0.9$, we expect to be close to the apex of intensity of the
electronic correlations. For such a filling, the occupied (with respect
to $\sigma_{cc}^{\left(m\right)}\left(\mathbf{k}\right)$: Fig.\ \ref{fig:Eksk}
(top-central panel)) region in energy-momentum space across the three
COM solutions is instead quite different, although some similarities
can still be found. In particular, as regards the regions close to
the chemical potential at the $\Gamma$ point and along the main anti-diagonal
(the $X-Y$ line). The behavior of $n^{\left(m\right)}$ (Fig.\ \ref{fig:Eksk}
(top-right panel)) at $n=0.9$ and, in general, at intermediate and
large fillings, clearly shows that COM(3p) CB, which was the main
actor at low fillings, tends to systematically lose occupation in
favor of the LHB. Close to half filling, this latter eventually exceeds
the former in occupation and collects more and more of it on increasing
$U$ (not shown) while the CB depletes on approaching the metal-insulator
transition. As regards COM(2p,$p>0$) instead, UHB plays a minor role
all the way up to the metal-insulator transition. It collects a small
fraction of the electronic occupation and only above a certain intermediate
value of the filling. Moving to the double occupancy $D$, it is worth
noticing that the interested regions in energy-momentum space are
quite different. COM(3p) receives significant contributions from both
the LHB, close to the $M$ point and along the main anti-diagonal
(the $X-Y$ line), and the CB, along the main anti-diagonal (the $X-Y$
line). COM(2p,$p<0$) only from the LHB close to the $M$ point. COM(2p,$p>0$)
almost only from the UHB close to the $\Gamma$ point. It is now clear
why the behavior of the double occupancy among the three COM solutions
is very similar (almost identical) at small fillings between COM(3p)
and COM(2p,$p<0$) - it comes from particles residing in the very
same region in energy-momentum space - and only accidentally similar
between COM(3p) and COM(2p,$p>0$) at large fillings.

\begin{figure*}
\begin{centering}
\begin{tabular}{cc}
\includegraphics[width=0.5\textwidth]{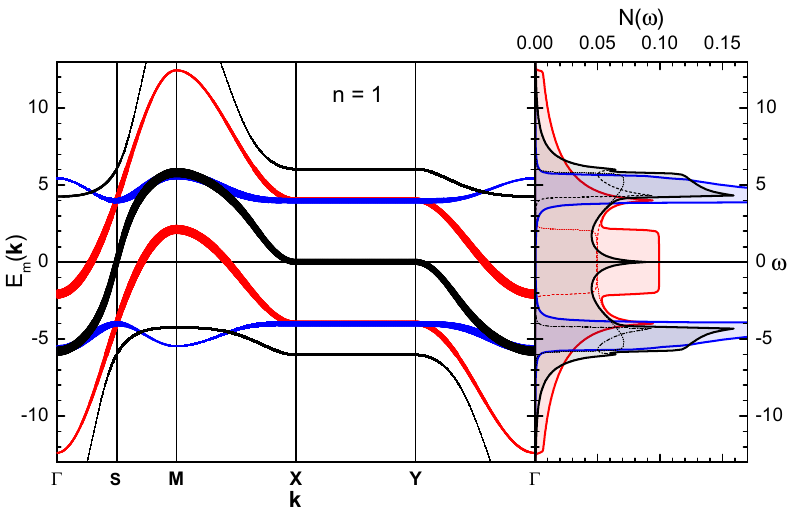} & \includegraphics[width=0.5\textwidth]{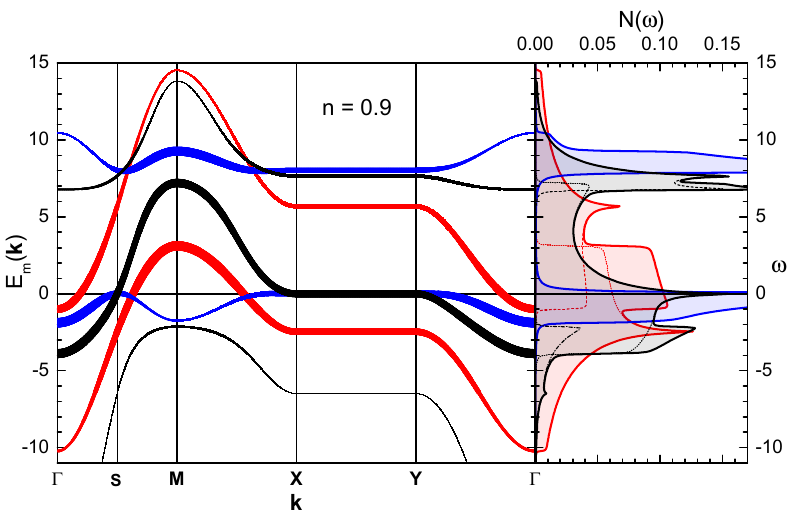}\tabularnewline
\multicolumn{2}{c}{\includegraphics[width=0.75\textwidth]{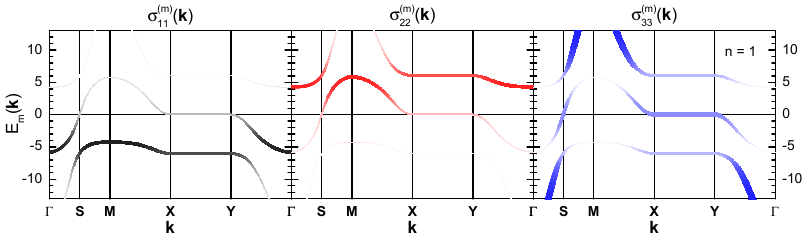}}\tabularnewline
\multicolumn{2}{c}{\includegraphics[width=0.75\textwidth]{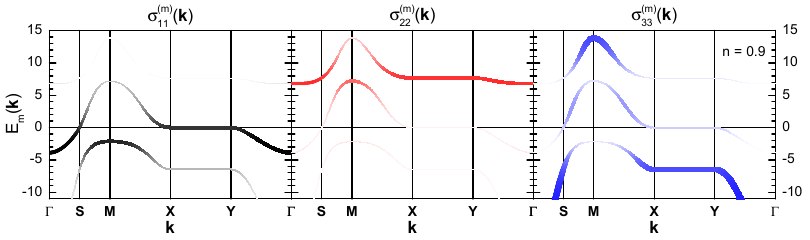}}\tabularnewline
\end{tabular}
\par\end{centering}

\caption{(top) Energy bands $E_{m}\left(\mathbf{k}\right)$ along the principal
directions of the first Brillouin zone ($\Gamma=(0,0)$ $\to$ $S=(\nicefrac{\pi}{2},\nicefrac{\pi}{2})$
$\to$ $M=(\pi,\pi)$ $\to$ $X=(\pi,0)$ $\to$ $Y=(0,\pi)$ $\to$
$\Gamma=(0,0)$) at $T\cong0$, $U=8$ and two different values of
the filling $n=1$ (left) and $n=0.9$ (right) for COM(3p) (variable-thickness
black line), COM(2p,$p>0$) (variable-thickness red line) and COM(2p,$p<0$)
(variable-thickness blue line). The thickness of each band is proportional
to the value of the corresponding electronic spectral density weight
$\sigma_{cc}^{\left(m\right)}\left(\mathbf{k}\right)$. The corresponding
densities of states $N\left(\omega\right)$ are reported using thick
lines and shaded areas of the same colors: thin dashed, dotted and
dot-dashed lines marks the contributions of the various bands. (central-bottom)
Energy bands $E_{m}\left(\mathbf{k}\right)$ along the principal directions
of the first Brillouin zone ($\Gamma=(0,0)$ $\to$ $S=(\nicefrac{\pi}{2},\nicefrac{\pi}{2})$
$\to$ $M=(\pi,\pi)$ $\to$ $X=(\pi,0)$ $\to$ $Y=(0,\pi)$ $\to$
$\Gamma=(0,0)$) at $T\cong0$, $U=8$ and two different values of
the filling $n=1$ (central) and $n=0.9$ (bottom) for COM(3p). The
thickness of each band is proportional to the value of the corresponding
spectral density weight (left) $\sigma_{11}^{\left(m\right)}\left(\mathbf{k}\right)$,
(central) $\sigma_{22}^{\left(m\right)}\left(\mathbf{k}\right)$ and
(right) $\sigma_{33}^{\left(m\right)}\left(\mathbf{k}\right)$.\label{fig:EkskDOS}}
\end{figure*}

In Fig.\ \ref{fig:EkskDOS} (top panels), we report the energy bands
$E_{m}\left(\mathbf{k}\right)$ along the principal directions of
the first Brillouin zone ($\Gamma=(0,0)$ $\to$ $S=(\nicefrac{\pi}{2},\nicefrac{\pi}{2})$
$\to$ $M=(\pi,\pi)$ $\to$ $X=(\pi,0)$ $\to$ $Y=(0,\pi)$ $\to$
$\Gamma=(0,0)$) at $T\cong0$, $U=8$ and two different values of
the filling $n=1$ and $n=0.9$. The thickness of each band is proportional
to the value of the corresponding electronic spectral density weight
$\sigma_{cc}^{\left(m\right)}\left(\mathbf{k}\right)$. The corresponding
density of states, $N\left(\omega\right)=\frac{1}{N}\sum_{\mathbf{k}}\left\{ -\frac{1}{\pi}\Im\left[G_{cc}\left(\mathbf{k},\omega\right)\right]\right\} $,
is also reported. The latter depends on both the electronic spectral
weight, $\sigma_{cc}^{\left(m\right)}\left(\mathbf{k}\right)$, and
the \emph{effective} velocity in the band $m$, $\mathbf{v}_{m}\left(\mathbf{k}\right)=\nabla_{\mathbf{k}}E_{m}\left(\mathbf{k}\right)$,
\begin{align}
N\left(\omega\right) & =\frac{1}{N}\sum_{\mathbf{k}}\sum_{m}\sigma_{cc}^{\left(m\right)}\left(\mathbf{k}\right)\delta\left(\omega-E_{m}\left(\mathbf{k}\right)\right)\nonumber \\
 & =\sum_{p}\left\Vert \mathbf{v}_{m}\left(\mathbf{k}_{m}^{\left(p\right)}\left(\omega\right)\right)\right\Vert ^{-1}\sigma_{cc}^{\left(m\right)}\left(\mathbf{k}_{m}^{\left(p\right)}\left(\omega\right)\right)\label{eq:deltaEk}
\end{align}
where $\mathbf{k}_{m}^{\left(p\right)}\left(\omega\right)$ are the
zeros of $\omega-E_{m}\left(\mathbf{k}\right)=0$.

At $n=1$ (Fig.\ \ref{fig:EkskDOS} (top-left panel)), COM(3p) CB
is pinned to the chemical potential along the main anti-diagonal (the
$X-Y$ line). The corresponding van Hove singularity in the density
of states lies exactly at the Fermi level and the Luttinger theorem
is satisfied. Three is the minimal number of poles necessary to satisfy
the Luttinger theorem in the Hubbard model. The van Hove peak at the
chemical potential gets weaker and weaker on increasing $U$ (not
shown), completely disappears only for $U\rightarrow\infty$, and
its weight becomes almost negligible with respect to that in the Hubbard
subbands for values of U as large as 12, which is the critical value
for the metal-insulator transition in COM(2p,$p>0$). This signals
an absence of a net transition, but it also manifests a clear tendency
towards it.

This is the main drawback of having chosen as third field $c_{s}$
that is not an eigenoperator of the interacting term of the Hamiltonian
(\ref{eq:Ham}) and, consequently, does not interpret exactly the
scale of energy of $U$. Choosing the whole $\pi$ would not have
solved this, obviously. The introduction of $c_{s}$ as third field
in the operatorial basis improves enormously - up to making it practically
exact in many cases - the description of momentum-integrated quantities.
This clearly implies that the overall physical content of the chosen
third field is exactly what was needed to definitely improve the two-pole
solutions through a better description of the nearest-neighbor spin-spin
correlations (i.e. of the energy scale of $J=\nicefrac{4t^{2}}{U}$).
On the other hand, the analysis of COM(3p) bands shows that the CB
does not reflect correctly the energy scale of $U$ instead, at least
as regards its central portion. On increasing $U$, CB stretches out
(not shown) keeping its maximum (the $M$ point) at about $\nicefrac{U}{2}$
and its minimum (the $\Gamma$ point) at about $-\nicefrac{U}{2}$,
that is, within the UHB and the LHB, respectively, while the spectral
weight moves rapidly from the central portion, pinned at the chemical
potential, towards the extrema. The CB would like to split in two
- following its $\xi$ and $\eta$ components - and open up a gap
of the order $U$, but it never manages to do so up to $U\rightarrow\infty$
because of its $c$-like component, that is, of the component not
resolved in $U$ as in a mean-field treatment of the model.

This is also shown by the spectral weight decoration, according to
the three fields of the basis, of COM(3p) bands in Fig.\ \ref{fig:EkskDOS}
(central panel). It is worth reminding that $\sigma_{33}^{\left(m\right)}\left(\mathbf{k}\right)$
does not directly enter $\sigma_{cc}^{\left(m\right)}\left(\mathbf{k}\right)$,
yet it is the best measure of which regions in the energy-momentum
space are more affected by $c_{s}$. On one hand, the presence of
$c_{s}$ in the basis allows to access those states missing in the
two-pole description and resulting in an almost exact description
of many relevant quantities. On the other hand, the energy-momentum
relation/position of some of these states is simply wrong on the energy
scale of $U$. Integrating over momentum this is not so relevant,
but becomes evident resolving the bands of the system. As a matter
of fact, COM(3p) solution performs so well that is worth analyzing
it in detail in order to deeply understand its relevant ingredients
so to have an absolutely preferential starting point to improve upon
it as regards just this single issue. Along this line, it is very
remarkable that COM(3p) CB exactly coincide with COM(2p,$p<0$) LHB
and UHB at the $\Gamma$ and the $M$ points, respectively, as well
as COM(3p) LHB and UHB are very close (just concavity differs) to
COM(2p,$p<0$) LHB and UHB at the $M$ and the $\Gamma$ points, respectively.
It is rather evident that, as regards the physics of the lower and
upper Hubbard bands, that is, the physics at the scale of energy of
$U$, COM(3p) builds upon COM(2p,$p<0$). COM(2p,$p>0$) simply describes
a different physics and it is very difficult to compare the two solutions.
Looking now at the density of states, it is clear that COM(3p), as
well as COM(2p,$p<0$), features peaks in the LHB and in the UHB with
the expected strong reduction of the bandwidth from $8t$ to something
of the order $J$ according to the reduced mobility of the electrons
in a strongly correlated almost-antiferromagnetic background. This
is also reflected by the very strong asymmetry, in shape and occupation,
with respect to the main anti-diagonal in the LHB and in the UHB.

At $n=0.9$ (Fig.\ \ref{fig:EkskDOS} (top-right panel)), it is evident
that COM(3p) CB is still almost pinned to the chemical potential along
the main anti-diagonal (the $X-Y$ line); the van Hove singularity
lies little below the Fermi level. Accordingly, changing the filling
in this region of low doping (from $n=0.85$ to $n=1$) has mainly
the effect to induce a transfer of spectral weight between the bands
and between their components in terms of fields of the basis, as one
would expect in a strongly correlated regime, rather than shifting
the chemical potential more or less rigidly within the bands, as it
could be expected at small fillings and weak interactions. It is also
evident, looking at the density of states too, that the LHB has still
a minor role with respect to the CB, which collects the vast majority
of the occupied states, as also shown by the spectral weight decoration
of COM(3p) bands in Fig.\ \ref{fig:EkskDOS} (bottom panel). It is
worth noting that the spin-spin correlations are already present,
but not yet so strong to determine the reduction of the bandwidth
in the energy-momentum space region shared by CB and LHB. As well
as at $n=1$, although less because of the lack of particle-hole symmetry
that is ruling the physics at half filling, COM(3p) bands are quite
close to COM(2p,$p<0$) ones.

\subsection{Charge, spin and pair correlation functions\label{sec:corr}}

\begin{figure}
\begin{centering}
\begin{tabular}{c}
\includegraphics[width=8cm]{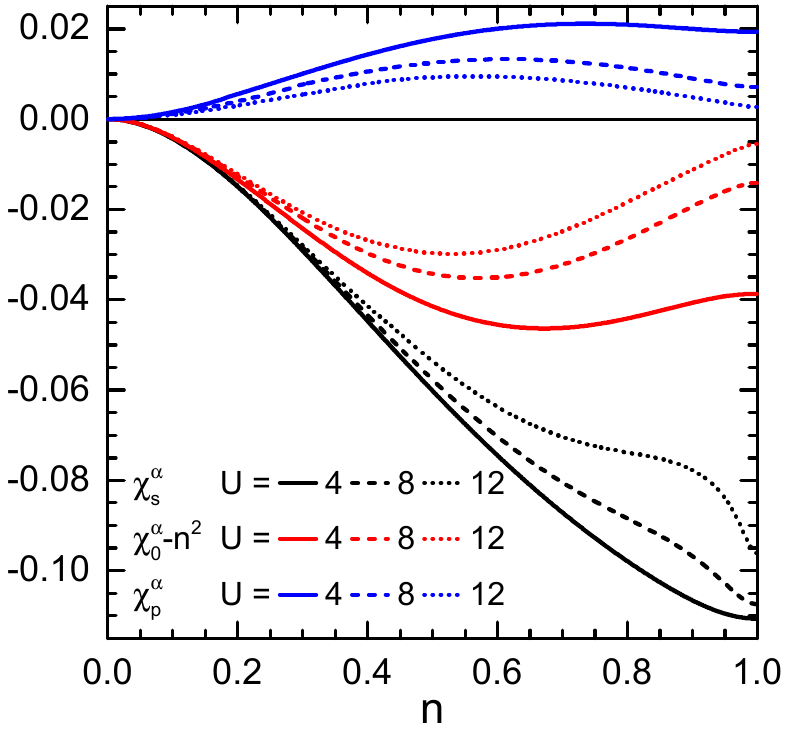}\tabularnewline
\includegraphics[width=8cm]{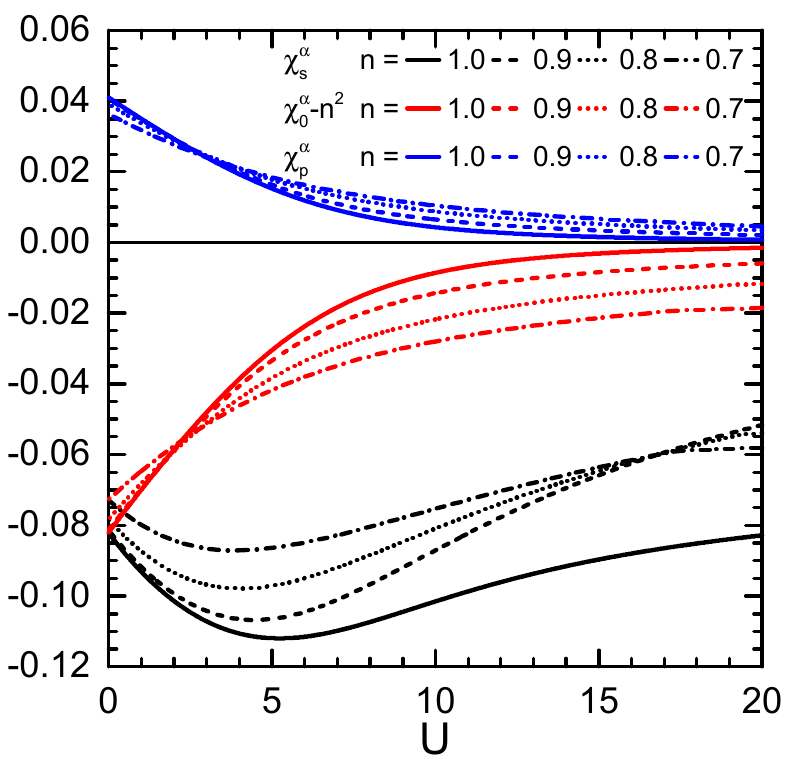}\tabularnewline
\end{tabular}
\par\end{centering}

\caption{Spin $\chi_{s}^{\alpha}$ (black lines), charge $\chi_{0}^{\alpha}$
(red lines) and pair $\chi_{p}^{\alpha}$ (blue lines) nearest-neighbor
correlation functions as functions of filling $n$ (top) and on-site
Coulomb repulsion $U$ (bottom) at $T\approxeq0$ for different values
of $U$ ((solid lines) $4$, (dashed lines) $8$ and (dotted lines)
$12$) and $n$ ((solid lines) $0.7$, (dashed lines) $0.8$, (dotted
lines) $0.9$ and (dot-dashed lines) $1$), respectively. Charge nearest-neighbor
correlation function $\chi_{0}^{\alpha}$ has been diminished of its
core non-fluctuating value $n^{2}$.\label{fig:chi}}
\end{figure}

In Fig.\ \ref{fig:chi}, we report the spin $\chi_{s}^{\alpha}$,
charge $\chi_{0}^{\alpha}$ and pair $\chi_{p}^{\alpha}$ nearest-neighbor
correlation functions as functions of filling $n$ and on-site Coulomb
repulsion $U$ at $T\approxeq0$ for different values of $U$ ($4$,
$8$ and $12$) and $n$ ($0.7$, $0.8$, $0.9$ and $1$), respectively.
Charge nearest-neighbor correlation function $\chi_{0}^{\alpha}$
has been diminished of its core non-fluctuating value $n^{2}$ in
order to put all three correlation functions on the same \emph{fluctuation}
footing and make possible a direct comparison of their values.

The first and more relevant consideration that can be drawn looking
at both panels at once regards the evident and net predominance of
the spin correlations over the charge and pair ones in the relevant
range of doping and on-site Coulomb repulsion. This justifies a posteriori
the choice of $c_{s}$ as third field in the basis instead of $\pi$
or of any other of its components (charge or pair).

Spin correlations monotonically increase with filling reaching their
maximum at half filling as expected. On approaching half filling,
an increasing on-site Coulomb repulsion has two main competing effects.
On one side, one needs a value of $U$ large enough to establish the
perturbative virtual process at the basis of the appearance of the
scale of energy of $J=\nicefrac{4t^{2}}{U}$. On the other side, $J$
is inversely proportional to $U$ and can simply vanish for high enough
values of this latter or just become too small to induce strong enough
antiferromagnetic correlations in presence of sufficiently high doping.
This occurrence has the obvious effect to frustrate the antiferromagnetic
order. Too high values of $U$ tend to forbid the virtual process
and to favor a ferromagnetic order at half filling instead of an antiferromagnetic
one. These facts can explain the quite strange behavior at $U=12$,
already somewhat visible at $U=8$. Spin correlations are weaker at
$U=12$ than at smaller values of $U$, in particular at large-intermediate
doping, then suddenly increase much faster for small enough doping.
At half filling, any not exceedingly small value of $J$ favors an
antiferromagnetic ordering. In order to reach a full understanding
of the reported results, it should be taken into account that we are
studying just the homogeneous paramagnetic phase and no real ordering
can be expected. Charge and pair correlations feature a maximum around
$n=0.5$, where a checkerboard charge order or a double-checkerboard
pair order could establish, and tend to vanish at half filling on
increasing the on-site Coulomb repulsion that quenches all their fluctuations.

As a function of the on-site Coulomb repulsion $U,$ $\chi_{s}^{\alpha}$
presents a behavior that perfectly agrees with the previous explanation.
For small increasing values of $U$, spin correlations also increase
following the systematic growth of the number of single-occupied sites.
For large increasing values of $U$, spin correlations decrease following
the systematic reduction of $J$. In between, $\chi_{s}^{\alpha}$
crosses its maximum at a critical value of $U$ that depends on the
value of the filling and on the homogenous-paramagnetic boundary conditions.
In fact, we can expect a higher critical value if true antiferromagnetic
order would be allowed. It can be inferred that, for small enough
doping, the spin correlations will turn positive (ferromagnetic) for
sufficiently high values of $U$. It is also evident that half filling
is quite special as the spin correlations remain stronger and antiferromagnetic
for much higher values of $U$. Charge and pair correlations, instead,
systematically decrease on reducing the doping and increasing the
on-site Coulomb repulsion showing how their fluctuations completely
quench on approaching the metal-insulator transition.

\section{Summary and Outlook\label{sec:Summary}}

We have presented and analyzed in detail a three-pole solution for
the two-dimensional Hubbard model within the Composite Operator method
framework. The third field, after the two Hubbard ones, has been chosen
according to the hierarchy of the equations of motion, but picking
up only the operatorial term related to spin correlations/fluctuations
in order to both simplify the calculations and highlight the most
relevant physics. It is worth noticing that charge and pair operatorial
terms have been just projected on the basis chosen and not simply
neglected. This choice is justified and promoted, a posteriori, by
the very good results obtained in the reported comprehensive comparison
with the data obtained by different numerical methods for momentum-integrated
quantities (e.g. local properties) as functions of all model parameters
(filling, on-site Coulomb repulsion and temperature) as well as for
the energy bands of the system. Spin correlations, as expected, are
also shown to be the most relevant in intensity and the richest in
features so to self-consistently validate the basis choice. The proposed
solution has shown to be able to catch many relevant features of the
numerical data: the crossover between weak and intermediate-strong
correlations on varying both filling and on-site Coulomb repulsion,
the way to approach the Mott-Hubbard metal-insulator transition, the
presence and energy position of shadow bands, and the exact shape
of the occupied portion of the dispersion. A comprehensive comparison
with other two-pole solutions has been also reported in order to better
understand the ultimate reasons of the very successful comparison
with the numerical data and to characterize the proposed solution
within the n-pole framework. On one hand, it is evident that the physical
content of the chosen third field is driving all improvements and
seems exhaustive as regards many relevant properties of the system
under analysis so to grant the right for, actually to call for the
need of, a comprehensive analysis to this solution. On the other hand,
it is also clear that the shape/behavior of the energy-momentum dispersion
close to the chemical potential at half filling dictated by the actual
choice of the third field suffers from this latter not being an eigenoperator
of the interacting term of the Hamiltonian. The field has the right
momentum-integrated physical content, but not the exact momentum-resolved
one. This can definitely be improved with some efforts in reconsidering
the algebra of the operatorial fields involved and we are actually
working on this. At any rate, this solution performs already very
well as regards many of the relevant properties of the analyzed model.
\begin{acknowledgments}
The author wishes to thank Massimo Capone, Peter Prelov\v{s}ek and Giorgio
Sangiovanni for providing him their numerical data and for the many
insightful discussions. The author also wishes to thank the referees
for the improvements on the discussion and the presentation of the
manuscript they have fostered.
\end{acknowledgments}
\appendix

\section{Operatorial Projection\label{app:op_proj}}

Any field operator $\phi\left(i\right)$ can be projected on a set
of fields operators $\left\{ \psi_{m}\left(i\right)\right\} $ featuring
$I_{\psi_{m}\psi_{p}}\left(\mathbf{i},\mathbf{j}\right)=\delta_{mp}\delta_{\mathbf{ij}}I_{\psi_{m}\psi_{m}}$
(e.g. $\left\{ \xi\left(i\right),\eta\left(i\right)\right\} $) through
the following relation
\begin{equation}
\phi\left(i\right)\cong\sum_{m}\sum_{\mathbf{j}}\frac{I_{\phi\psi_{m}}\left(\mathbf{i},\mathbf{j}\right)}{I_{\psi_{m}\psi_{m}}}\psi_{m}\left(\mathbf{j},t\right)\label{eq:proj}
\end{equation}

We can rewrite $\chi_{s}^{\alpha}$ and $f_{s}$ as follows (we extracted
first the component of $f_{s}$ involving only two sites: $-\frac{1}{2}C_{c\xi}^{\alpha}$)
\begin{align}
\chi_{s}^{\alpha} & =-\frac{1}{3}\Tr\left\langle \phi_{s}\left(i\right)\otimes c^{\dagger}\left(i\right)\right\rangle \label{eq:chias_proj}\\
f_{s} & =-\frac{1}{2}C_{c\xi}^{\alpha}-\frac{1}{6}\Tr\left\langle \phi_{s}\left(i\right)\otimes c^{\dagger\alpha\left(\beta\right)}\left(i\right)\right\rangle \nonumber \\
 & -\frac{1}{12}\Tr\left\langle \phi_{s}\left(i\right)\otimes c^{\dagger\alpha\left(\eta\right)}\left(i\right)\right\rangle \label{eq:fs_proj}
\end{align}
where $\phi_{s}\left(i\right)=n_{k}^{\alpha}\left(i\right)\sigma_{k}\cdot c\left(i\right)$.
$\left(\varrho^{\alpha}\left(i\right)\otimes\right)\varphi^{\alpha\left(\beta\right)}\left(i\right)$
and $\left(\varrho^{\alpha}\left(i\right)\otimes\right)\varphi^{\alpha\left(\eta\right)}\left(i\right)$
stand for an operator $\varphi$ sited on a site that is nearest-neigbor
($\alpha$) of site $\mathbf{i}$ and second-nearest-neighbor ($\beta$
and $\eta$, respectively) of the actual site where the operator $\varrho$
is sited (e.g. $\varrho^{\alpha}\left(i\right)\otimes\varphi^{\alpha\left(\beta\right)}\left(i\right):$
$\varrho\left(i\pm\hat{x}\right)\otimes\varphi\left(i\pm\hat{y}\right)$,
$\varrho^{\alpha}\left(i\right)\otimes\varphi^{\alpha\left(\eta\right)}\left(i\right):$
$\varrho\left(i\pm\hat{x}\right)\otimes\varphi\left(i\mp\hat{x}\right)$).

Choosing $\left\{ \xi\left(i\right),\eta\left(i\right)\right\} $
as set of fields operators $\left\{ \psi_{m}\left(i\right)\right\} $,
we have the following relevant relations
\begin{align}
I_{\phi_{s}\xi}\left(\mathbf{i},\mathbf{j}\right) & =\frac{3}{2}\delta_{\mathbf{ij}}\chi_{s}^{\alpha}+3\alpha_{\mathbf{ij}}C_{c\xi}^{\alpha}\label{eq:phisxi}\\
I_{\phi_{s}\eta}\left(\mathbf{i},\mathbf{j}\right) & =-\frac{3}{2}\delta_{\mathbf{ij}}\chi_{s}^{\alpha}+3\alpha_{\mathbf{ij}}C_{c\eta}^{\alpha}\label{eq:phiseta}
\end{align}
Accordingly, we have the following projection for the field $\phi_{s}\left(i\right)$
\begin{equation}
\phi_{s}\left(i\right)\cong\frac{3}{2}\frac{\chi_{s}^{\alpha}}{I_{11}}\xi\left(i\right)-\frac{3}{2}\frac{\chi_{s}^{\alpha}}{I_{22}}\eta\left(i\right)+3\frac{C_{c\xi}^{\alpha}}{I_{11}}\xi^{\alpha}\left(i\right)+3\frac{C_{c\eta}^{\alpha}}{I_{22}}\eta^{\alpha}\left(i\right)\label{eq:phis_proj}
\end{equation}
that leads to the following closed relation for $\chi_{s}^{\alpha}$
(leading to (\ref{eq:chias}))
\begin{equation}
\chi_{s}^{\alpha}\cong-\frac{\chi_{s}^{\alpha}}{I_{11}}C_{\xi\xi}+\frac{\chi_{s}^{\alpha}}{I_{22}}C_{\eta\eta}-2\frac{C_{c\xi}^{\alpha}}{I_{11}}C_{c\xi}^{\alpha}-2\frac{C_{c\eta}^{\alpha}}{I_{22}}C_{c\eta}^{\alpha}\label{eq:chias_clo}
\end{equation}
and to the expression of $f_{s}$ in the main text (\ref{eq:fs}).
To get this latter, we used the geometrical relation $C_{\phi\psi}^{\alpha^{2}}=\frac{1}{4}C_{\phi\psi}+\frac{1}{2}C_{\phi\psi}^{\beta}+\frac{1}{4}C_{\phi\psi}^{\eta}$.

In the very same way, we can rewrite $\chi_{0}^{\alpha}$ as follows
\begin{equation}
\chi_{0}^{\alpha}=2n-\Tr\left\langle \phi_{0}\left(i\right)\otimes c^{\dagger}\left(i\right)\right\rangle \label{eq:chia0_proj}
\end{equation}
where $\phi_{0}\left(i\right)=n^{\alpha}\left(i\right)c\left(i\right)$.
Using the very same set of fields operators, we have the following
relevant relations
\begin{align}
I_{\phi_{0}\xi}\left(\mathbf{i},\mathbf{j}\right) & =\delta_{\mathbf{ij}}\left(n-\frac{1}{2}\chi_{0}^{\alpha}\right)+\alpha_{\mathbf{ij}}C_{c\xi}^{\alpha}\label{eq:phi0xi}\\
I_{\phi_{0}\eta}\left(\mathbf{i},\mathbf{j}\right) & =\frac{1}{2}\delta_{\mathbf{ij}}\chi_{0}^{\alpha}+\alpha_{\mathbf{ij}}C_{c\eta}^{\alpha}\label{eq:phi0eta}
\end{align}
Accordingly, we have the following projection for the field $\phi_{0}\left(i\right)$
\begin{equation}
\phi_{0}\left(i\right)\cong\frac{n-\frac{1}{2}\chi_{0}^{\alpha}}{I_{11}}\xi\left(i\right)+\frac{1}{2}\frac{\chi_{0}^{\alpha}}{I_{22}}\eta\left(i\right)+\frac{C_{c\xi}^{\alpha}}{I_{11}}\xi^{\alpha}\left(i\right)+\frac{C_{c\eta}^{\alpha}}{I_{22}}\eta^{\alpha}\left(i\right)\label{eq:phi0_proj}
\end{equation}
that leads to the following closed relation for $\chi_{0}^{\alpha}$
(leading to (\ref{eq:chia0}))
\begin{equation}
\chi_{0}^{\alpha}\cong2n-\frac{2n-\chi_{0}^{\alpha}}{I_{11}}C_{\xi\xi}-\frac{\chi_{0}^{\alpha}}{I_{22}}C_{\eta\eta}-2\frac{C_{c\xi}^{\alpha}}{I_{11}}C_{c\xi}^{\alpha}-2\frac{C_{c\eta}^{\alpha}}{I_{22}}C_{c\eta}^{\alpha}\label{eq:chia0_clo}
\end{equation}

\[
\]
Once more, we can rewrite $\chi_{p}^{\alpha}$ as follows
\begin{equation}
\chi_{p}^{\alpha}=\left\langle \phi_{p}\left(i\right)c_{\uparrow}^{\dagger}\left(i\right)\right\rangle \label{eq:chiap_proj}
\end{equation}
where $\phi_{p}\left(i\right)=\left[c_{\uparrow}\left(i\right)c_{\downarrow}\left(i\right)\right]^{\alpha}c_{\downarrow}^{\dagger}\left(i\right)$.
Using the very same set of fields operators, we have the following
relevant relations
\begin{align}
I_{\phi_{p}\xi}\left(\mathbf{i},\mathbf{j}\right) & =\delta_{\mathbf{ij}}\chi_{p}^{\alpha}+\alpha_{\mathbf{ij}}C_{c\eta}^{\alpha}\label{eq:phipxi}\\
I_{\phi_{p}\eta}\left(\mathbf{i},\mathbf{j}\right) & =-\delta_{\mathbf{ij}}\chi_{p}^{\alpha}+\alpha_{\mathbf{ij}}C_{c\xi}^{\alpha}\label{eq:phipeta}
\end{align}
Accordingly, we have the following projection for the field $\phi_{p0}\left(i\right)$
\begin{equation}
\phi_{p}\left(i\right)\cong\frac{\chi_{p}^{\alpha}}{I_{11}}\xi\left(i\right)-\frac{\chi_{p}^{\alpha}}{I_{22}}\eta\left(i\right)+\frac{C_{c\eta}^{\alpha}}{I_{11}}\xi^{\alpha}\left(i\right)+\frac{C_{c\xi}^{\alpha}}{I_{22}}\eta^{\alpha}\left(i\right)\label{eq:phip_proj}
\end{equation}
that leads to the following closed relation for $\chi_{p}^{\alpha}$
(leading to (\ref{eq:chiap}))
\begin{equation}
\chi_{p}^{\alpha}\cong\frac{\chi_{p}^{\alpha}}{I_{11}}C_{\xi\xi}-\frac{\chi_{p}^{\alpha}}{I_{22}}C_{\eta\eta}+\frac{C_{c\eta}^{\alpha}}{I_{11}}C_{c\xi}^{\alpha}+\frac{C_{c\xi}^{\alpha}}{I_{22}}C_{c\eta}^{\alpha}\label{eq:chiap_clo}
\end{equation}


\begin{thebibliography}{118}
\bibitem[{eig()}]{eigen}
\bibinfo{note}{An eigenoperator $\psi$ of a Hamiltonian $H$, or of one or more
  terms of this latter, is an operator whose current $J$ with respect to $H$,
  or to one or more terms of this latter, is proportional to the operator
  itself: $J=\mathrm{i}\frac{\partial}{\partial
  t}\psi=\left[\psi,H\right]=\varepsilon\psi$. The proportionality constant
  $\varepsilon$ is named eigenenergy of the eigenoperator $\psi$ with respect
  to $H$, or to one or more terms of this latter.}
\bibitem[{vec()}]{vectorial}
\bibinfo{note}{This vector is of the type $v_{l,\sigma}$ where the index $l$
  individuates an operator in the operatorial basis and the index $\sigma$ is
  the spin indices related to the spinorial notation. The summation over these
  two indexes is usually understood. See also
  Ref.~\protect\onlinecite{matricial}.}
\bibitem[{mat()}]{matricial}
\bibinfo{note}{This matrix is of the type $A_{l,m;\sigma,\sigma'}$ where the
  indexes $l$ and $m$ individuate an operator in the operatorial basis and the
  indexes $\sigma$ and $\sigma'$ are the spin indices related to the spinorial
  notation. The summation over these four indexes is usually understood. For
  instance:
  $J\left(i\right)=\sum_{\mathbf{j}}\varepsilon\left(\mathbf{i},\mathbf{j}\right)\psi\left(\mathbf{j},t\right)+\delta
  J(i)$ $\Rightarrow$
  $J_{l,\sigma}(i)=\sum_{m;\sigma';\mathbf{j}}\varepsilon_{l,m;\sigma,\sigma'}(\mathbf{i},\mathbf{j})\psi_{m,\sigma'}\left(\mathbf{j},t\right)+\delta
  J_{l,\sigma}(i)$, $G(i,j)=\left\langle
  \mathcal{R}\left[\psi(i)\psi^{\dagger}(j)\right]\right\rangle$ $\Rightarrow$
  $G_{l,m;\sigma,\sigma'}(i,j)=\left\langle
  \mathcal{R}\left[\psi_{l,\sigma}(i)\psi_{m,\sigma'}^{\dagger}(j)\right]\right\rangle$.}
\bibitem[{sem()}]{semi-positive}
\bibinfo{note}{The product of two symmetric matrices has real eigenvalues if
  one of the two is semi-positive, that is, it has positive or null
  eigenvalues.}
\end{thebibliography}

\end{document}